\documentclass[prd,nofootinbib,preprint,superscriptaddress,preprintnumbers,floatfix,11pt]{revtex4-1}
\pdfoutput=1
\usepackage{color}
\usepackage{graphicx}
\usepackage{tabularx}
\usepackage{xspace}

\usepackage{float}

\usepackage{verbatim}
\usepackage{amsmath}
\usepackage{amssymb}
\usepackage{url}
\usepackage{bbold}
\usepackage{slashed}
\usepackage{array}

\usepackage{multirow}
\usepackage{threeparttable}
\usepackage{paralist}
\usepackage{xspace}
\usepackage{upgreek}
\usepackage{lipsum}

\newcommand{\avg}[1]{\left< #1 \right>} 

\newcommand{\abs}[1]{\left| #1 \right|} 

\newcommand\colvec[3][]{\begin{pmatrix}\ifx\relax#1\relax\else#1\\\fi#2\\#3\end{pmatrix}}

\definecolor{darkmagenta}{rgb}{0.55, 0.0, 0.55}

\newcommand{\beq}{\begin{equation}}
\newcommand{\beqn}{\begin{eqnarray}}
\newcommand{\eeq}{\end{equation}}
\newcommand{\eeqn}{\end{eqnarray}}
\newcommand\numberthis{\addtocounter{equation}{1}\tag{\theequation}}
\newcommand{\dbar}{\ensuremath{\mathchar'26\mkern-12mu d}}

\newcommand{\MeV}{ \textrm{MeV} }

\definecolor{blue}{rgb}{0.1,0.4,0.6}
\definecolor{navy}{rgb}{0.1,0.2,0.5}

\usepackage[colorlinks]{hyperref}
\hypersetup{
     colorlinks   = true,
     citecolor    = blue,
	linkcolor = navy
}

\begin{document}

\title{Making dark matter out of light: freeze-in from plasma effects}
\author{Cora Dvorkin}
\email{cdvorkin@g.harvard.edu}
\affiliation{Department of Physics, Harvard University, Cambridge, MA 02138, USA}
\author{Tongyan Lin}
\email{tongyan@physics.ucsd.edu}
\affiliation{Department of Physics, University of California, San Diego, CA 92093, USA}
\author{\& Katelin Schutz}
\email{kschutz@berkeley.edu}
\affiliation{Berkeley Center for Theoretical Physics, University of California, Berkeley, CA 94720, USA}

\date{\today}

\begin{abstract} \noindent
Dark matter (DM) could couple to particles in the Standard Model (SM) through a light vector mediator. In the limit of small coupling, this portal could be responsible for producing the observed DM abundance through a mechanism known as freeze-in. Furthermore, the requisite DM-SM couplings provide a concrete benchmark for direct and indirect searches for DM.
In this paper, we present updated calculations of the relic abundance for DM produced by freeze-in through a light vector mediator. We identify an additional production channel: the decay of photons that acquire an in-medium plasma mass. These plasmon decays are a dominant channel for DM production for sub-MeV DM masses, and including this channel leads to a significant reduction in the predicted signal strength for DM searches. Accounting for production from both plasmon decays and annihilations of SM fermions, the DM acquires a highly non-thermal phase space distribution which impacts the cosmology at later times; these cosmological effects will be explored in a companion paper.
\end{abstract}

\maketitle

\section{Introduction}

One of the most well-studied mechanisms for setting the observed dark matter (DM) abundance is thermal freeze-out, where DM is in equilibrium with the Standard Model (SM) thermal bath at very early times. The DM abundance is then depleted through annihilations at later times until the DM drops out of chemical equilibrium. The appeal of this mechanism is that the final relic abundance is generally independent of the high-temperature initial conditions at reheating. Furthermore, producing the observed relic abundance requires a particular thermally averaged annihilation cross section in most thermal freeze-out scenarios, $\langle \sigma v\rangle \sim 10^{-26}$ cm$^3$/s. This weak-scale cross section provides a target that can be probed by direct and indirect detection experiments. Assuming the relic abundance is set by annihilations to SM particles, then consistency with Big Bang Nucleosynthesis (BBN) generally requires that thermal freeze-out candidates have masses $m_\chi \gtrsim 1$~MeV~\cite{Boehm:2013jpa, Nollett:2013pwa,Nollett:2014lwa}. 
The appealing simplicity of this scenario has led to an enormous number of DM searches targeting the thermal freeze-out mechanism, with a particular emphasis on weakly interacting massive particle (WIMP) candidates in the $m_\chi\sim$~GeV$-$TeV mass range. More recently, there has been a growing interest in $m_\chi\sim$~MeV$-$GeV thermal candidates where interactions with the SM or within a hidden sector deplete the DM density to the observed value \cite{Boehm:2003hm,Pospelov:2007mp,Feng:2008ya,Hochberg:2014dra,Hochberg:2014kqa,Hochberg:2018rjs,Choi:2017zww,DAgnolo:2018wcn,DAgnolo:2017dbv,DAgnolo:2015ujb,Pappadopulo:2016pkp,Cline:2017tka,Kopp:2016yji}. 

The freeze-in mechanism for DM production is a compelling alternative to thermal freeze-out, where DM is instead produced by feeble, sub-Hubble interactions of SM particles~\cite{Asaka:2005cn,Asaka:2006fs,Gopalakrishna:2006kr,Page:2007sh,Hall:2009bx,Bernal:2017kxu}. If the dominant freeze-in process is annihilation of SM particles into DM via a light mediator, then many of the appealing features of thermal freeze-out are maintained. For annihilation through a mediator lighter than the DM, the thermal cross section typically scales as $\langle \sigma v \rangle \sim g_\chi^2 g_{\rm SM}^2/(4 \pi T)^2$ where $g_\chi$ is the mediator-DM coupling, $g_{\rm SM}$ is the mediator-SM coupling, and $T$ is the SM temperature. With this scaling, DM freeze-in dominantly occurs at the lowest temperature where the process is kinematically accessible, and thus the mechanism is not sensitive to the reheat scale.\footnote{We assume the minimal scenario where the dark sector is not populated in abundance at reheating.}

Freeze-in through a light vector mediator has emerged as a key benchmark for sub-GeV direct detection experiments. Producing the observed DM relic abundance implies a tiny value for the coupling constants, which is difficult to target with accelerator searches. However, sufficiently light mediators give rise to scattering cross sections that scale as $\sigma \propto 1/v^4$ for relative velocity $v$, implying that the kinematics of the Milky Way (where $v\sim 10^{-3}$) can enhance the detectability of DM coupling to a light mediator. If the mediator also couples to charged SM fermions, then the DM can scatter off of electrons or nuclei and may be detectable with the next generations of direct detection experiments~\cite{Essig:2011nj,Essig:2012yx,Essig:2015cda,Hochberg:2016ntt,Derenzo:2016fse,Hochberg:2017wce,Knapen:2017ekk,Griffin:2018bjn,Schutz:2016tid,Knapen:2016cue,Hochberg:2015pha,Hochberg:2015fth} (see also Ref.~\cite{Battaglieri:2017aum} for a recent review). Indeed, recent experimental results by SENSEI~\cite{Crisler:2018gci,sensei}, SuperCDMS~\cite{Agnese:2018col}, and DarkSide~\cite{Agnes:2018oej} are demonstrating significant progress towards achieving the sensitivity needed in the MeV-GeV mass range. It was also shown recently that Xenon1T~\cite{Aprile:2018dbl} is for the first time constraining freeze-in in the GeV-TeV mass range~\cite{Hambye:2018dpi}. 

In the keV$-$MeV DM mass range, freeze-in is the leading scenario that could be tested by proposed low-threshold direct detection experiments. Refs.~\cite{Knapen:2017xzo,Green:2017ybv} studied the possible direct detection cross sections in models of sub-MeV DM, finding that it would be difficult to observe thermal freeze-out scenarios (even purely within a dark sector) due to a combination of BBN, CMB, fifth force, and stellar emission constraints. Obtaining accurate predictions of freeze-in is thus an important step in the program to search for low-mass DM. While freeze-in from electron-positron annihilations via a light vector mediator has been studied in the past~\cite{Chu:2011be,Essig:2011nj}, in this work we thoroughly explore a previously overlooked production mechanism: freeze-in through plasma effects. The contribution of plasma effects to dark sector thermalization was estimated earlier in Refs.~\cite{Davidson:2000hf,Vogel:2013raa} and the effect on freeze-in via a heavy mediator was recently considered in Ref.~\cite{An:2018nvz} as we were in the late stages of completing this work, but it was not included in previous studies of freeze-in through a light vector mediator. We find that the plasma production of DM is a dominant channel for sub-MeV DM masses, and will therefore restrict our discussion to this mass range. The additional contribution to the relic abundance implies that the target cross section for direct detection is lower by roughly an order of magnitude for the lowest experimentally accessible DM masses.

The rest of this paper is organized as follows. We begin in Section~\ref{sec:model} by reviewing the arguments for the simplest viable freeze-in models in the keV-MeV mass range: either pure millicharged DM arising from a DM hypercharge or \emph{effectively} millicharged DM that is coupled to an ultralight dark photon mediator. These two scenarios are almost phenomenologically identical, with the key difference being that DM-DM scattering can be parametrically larger when dark photon interactions are present. These DM candidates have recently received considerable attention in the context of the anomalous 21~cm global signal~\cite{Bowman:2018yin,barkana2018possible,Barkana:2018cct,Berlin:2018sjs,Munoz:2018pzp}. In Section~\ref{sec:prod} we compute the DM relic abundance from freeze-in via a light mediator. We include the effects of plasmon decays for the first time, and show the impact for direct detection. We then present the calculation of the phase space distribution for freeze-in DM in Section~\ref{sec:phasespace}. A summary of our results can be found in Section~\ref{summary}.
In a companion paper~\cite{inprep}, we  will apply the calculations of the phase-space distribution to cosmological observables, showing that the cosmic microwave background (CMB) and probes of large-scale structure (LSS) provide a strong complementary test of DM freeze-in for $m_\chi\sim$~keV$-$MeV. In particular, we find that existing cosmological constraints restrict $m_\chi \gtrsim$ tens of keV for freeze-in via a light mediator, and it will be possible to probe even higher masses with planned experiments. 

\section{Models for sub-MeV freeze-in \label{sec:model}}

\subsection{The case for light vector mediators}
\label{vectors}
The simplest observationally viable models for sub-MeV freeze-in through a light mediator can be divided into two classes, where (1) the DM only has interactions mediated by the SM photon or (2) the DM has interactions with an ultralight kinetically mixed dark photon. We note that models of millicharged DM~\cite{Davidson:2000hf,Dubovsky:2003yn} can fall under either category: they can arise as a limit of the dark-photon model where the dark photon is nearly massless, or they could be present as Dirac fermions with a tiny hypercharge.\footnote{Other models that have been considered in the past require giving neutrinos small charges as well~\cite{Foot:1992ui}, which we do not consider further due to strong experimental bounds on neutrino charge~\cite{Chen:2014dsa}.}

For sub-MeV freeze-in to be relevant for direct detection, vector mediators are the only observationally viable option due to stringent constraints on other light mediators with the requisite couplings to the SM, as outlined below. For direct detection of freeze-in, the mediator masses must be sufficiently small compared to the typical momentum transfer for scattering processes. If the mediators are heavier, then they do not give rise to the $v^{-4}$ enhancement that would render extremely feeble DM-SM interactions detectable on Earth. For nuclear recoils the relevant momentum scale is set by galactic kinematics $q \sim m_\chi v \sim 10^{-3} m_\chi$, while for electron recoils the typical electron momentum in the target material is most relevant $q \sim \alpha m_e \approx 4$~keV, where $m_e$ is the electron mass and $\alpha$ is the electromagnetic fine structure constant. Thus for sub-MeV DM, the experimentally relevant mediators have masses below $\mathcal{O}(1)$~keV.  

Assuming an annihilation cross section of SM fermions into DM with the form $\langle \sigma v\rangle \sim g_\chi^2 g_{\rm SM}^2/(4 \pi T)^2$, the relic abundance can be estimated as 
\begin{align}
    Y_\chi = \frac{n_\chi}{s} \sim  \frac{n_{\rm SM}^2 \langle \sigma v\rangle}{s H} \sim 2\times10^{-4} \, \frac{g_\chi^2 g_{\rm SM}^2 M_{\rm Pl} }{ T},
\end{align}
where $M_{\rm Pl}=1/\sqrt{8 \pi G}$ is the reduced Planck mass and we assumed $T \sim $ MeV. Then for $m_\chi\sim$ MeV, we find that $g_\chi g_{\rm SM} \simeq 10^{-12}$ to saturate the relic abundance. This order-of-magnitude estimate is in agreement with more detailed calculations below. Since obtaining the relic abundance from freeze-in requires $g_\chi g_{\rm SM} \sim 10^{-12}$, $g_{\rm SM}$ must be greater than $10^{-12}$ if we require the dark sector to be perturbative (i.e. $g_\chi~\lesssim~1$). Weakly coupled, sub-keV mediators can be emitted in stars, affecting their luminosity and lifetime. The observed properties of stars lead to strong bounds on such mediators, which we summarize here (see also  Refs.~\cite{Knapen:2017xzo,Green:2017ybv} where these bounds are collected and discussed in the context of sub-MeV DM models): 
\begin{itemize}
\item {\emph{Scalars and pseudoscalars coupled to electrons}} $-$ The strongest bound on a light scalar with interaction $g_{\phi ee} \phi \bar e e$ comes from helium ignition in red giants, with $g_{\phi e e} \lesssim 7 \times 10^{-16}$ for sub-keV masses~\cite{Hardy:2016kme}. For a sub-keV pseudoscalar, observations of white dwarfs give typical constraints of $g_{a e e} \lesssim 2 \times 10^{-13}$~\cite{Raffelt:2006cw,Viaux:2013lha,Bertolami:2014wua}. A caveat for most stellar emission bounds is that when the coupling is increased, the new particle may be trapped within the star and would not lead to anomalous energy loss. However, this would still affect energy transport in the star, which can be constrained for the range of couplings relevant for freeze-in through this mediator~\cite{Carlson:1988jg,Raffelt:1988rx}.
\item {\emph{Scalars and pseudoscalars coupled to nucleons}} $-$ Similar to the case of mediators coupling to electrons, red giants constrain $g_{\phi n n} \lesssim 10^{-12}$ for a scalar~\cite{Hardy:2016kme} and  $g_{a n n} \lesssim {\rm few} \times 10^{-10}$ for a pseudoscalar~\cite{Raffelt:2006cw,Giannotti:2017hny}. While the latter coupling appears at face value to be sufficiently large, freeze-in through baryons is largely suppressed after the QCD phase transition due to the low baryon number density. Therefore, in this case our estimate for the minimum $g_{\rm SM}$ with $T\sim 1$~MeV is much too low and freeze-in would have to occur with a larger value of $g_{\rm SM}$ that is in tension with stellar bounds. 
\item {\emph{Scalar mixing with the Higgs}} $-$ The bounds here are similar to those in the two previous cases, and it has been shown in Ref.~\cite{Krnjaic:2017tio} that freeze-in through this portal is only a viable mechanism for producing all of the DM for DM masses above a few hundred MeV. 
\item {\emph{Kinetically mixed dark photon}} $-$ In this case, the stellar constraints on $g_{\rm SM}$ decrease linearly with the mediator mass for masses below $\sim 100$ eV~\cite{An:2013yfc,An:2013yua} because of the in-medium plasma mass suppression of producing dark photons from SM interactions, as detailed in Eq.~\eqref{eq:inmedium_L} and the surrounding discussion in Section~\ref{sec:darkphoton}. From the collected bounds on dark photons from Refs.~\cite{Jaeckel:2013ija}, a dark photon can have $g_{\rm SM} > 10^{-12}$ when its mass is well below $1$~eV. At even lower masses, the coupling could be $\sim 10^{-3}$ for masses below $\lesssim 10^{-14}$~eV. 
\item {\emph{$B-L$ vector}} $-$ Stellar constraints on a $B-L$ vector are similar to that for the dark photon. However, for eV-scale and lighter mediator masses, a $B-L$ vector is also strongly constrained by fifth force searches (e.g.~\cite{Murata:2014nra,Adelberger:2003zx}), which limits the mediator-SM coupling to below $10^{-12}$. 
\end{itemize} 
Since we are focusing on the simplest benchmarks for direct detection, we do not consider more exotic possibilities with additional particles and interactions. From the bounds on new particles with the couplings described above, we conclude that freeze-in through a light mediator is viable either when the mediator is (1) the SM photon, and the DM has a tiny electric charge, or (2) when the mediator is an ultralight kinetically mixed dark photon. 

We discuss these two closely related scenarios in the rest of the section. In both cases, DM has an effective charge $Q e$ (or millicharge $Q$) with respect to the SM photon. This parameter determines the relic abundance, irrespective of which of the two models is under consideration. Both models allow for heat and momentum transfer between SM particles and DM during epochs when the typical relative velocities are low (as discussed in Section~\ref{sec:scattering}), which is relevant to observations of the CMB~\cite{Dvorkin:2013cea,Xu:2018efh,Slatyer:2018aqg,Kovetz:2018zan,Boddy:2018wzy,Dubovsky:2001tr,Boddy:2018kfv} and the cosmological 21~cm global signal~\cite{Bowman:2018yin,barkana2018possible,Barkana:2018cct,Berlin:2018sjs,Munoz:2018pzp}. The main phenomenological difference between these two possibilities is that DM-DM scattering via a dark photon can be parametrically larger than DM-DM scattering mediated by the SM photon, as discussed below. If present at a sufficient level, the DM self-scattering can play an important role in determining the DM phase space distribution at late times, well after freeze-in.

\subsection{DM with photon-mediated interactions \label{sec:millicharge}}
 
If the DM is a Dirac fermion $\chi$ with a tiny hypercharge $Q_Y$ (the only gauge-invariant, renormalizable operator leading to a bare millicharge), then it can interact via the SM photon. After electroweak symmetry breaking, the DM obtains an electric charge given by $e Q_Y \equiv eQ$ (taking the convention where the Gell-Mann Nishijima formula reads $Q = I_3 +Y$). Although there are also $Z$-mediated DM interactions, they are negligible for the relevant epochs where $T \ll m_Z$. This gives the simplest model of millicharged DM. It is difficult to incorporate such matter content into a Grand Unified Theory (GUT)~\cite{Okun:1983vw}; however, this scenario is economical in that it requires that no additional particles be introduced to the SM aside from the DM itself. 

The possibility that this DM candidate obtains its relic abundance by thermal freeze-out has been considered before in Ref.~\cite{McDermott:2010pa}, where it was shown to be excluded by structure formation when all of the DM is produced this way. Thus, freeze-in is the simplest remaining possibility for producing this DM candidate, with $g_\chi = eQ$ and $g_{\rm SM} = e$ in the language of the previous subsections.

There are stellar emission bounds on this DM candidate because the DM can be pair produced by the decay of plasmons in stars, leading to additional energy loss. These bounds are shown as the shaded region in our summary plot, Fig.~\ref{fig:summaryplot}. Constraints on DM pair produced in SN1987a were derived in Refs.~\cite{Davidson:2000hf,Chang:2018rso} and require $Q\lesssim 10^{-9}$ for $m_\chi$ up to a few MeV, which does not impact freeze-in. However, there are constraints for $m_\chi$ below $\mathcal{O}(10)$~keV from emission in white dwarfs, horizontal branch stars, and red giants (see Appendix of Ref.~\cite{Vogel:2013raa}). Note that the range of $m_\chi$ where stellar emission can constrain freeze-in is exponentially sensitive to assumptions about temperatures within the stars. In addition, the bounds derived are applicable in the weak coupling limit where the DM escapes cleanly from the star. For sufficiently large $Q$, DM emission could contribute to energy transport within the star and the effects have not been carefully studied in this regime. The couplings for freeze-in are large enough that they could be in this regime and stellar bounds on freeze-in should be regarded with care. 

The relevant interactions for the relic abundance and phase space distribution in this model are SM annihilations and plasma decay into the DM. DM-SM scattering can become important at late times but, as we discuss in Section~\ref{sec:scattering}, the effect must be small to be consistent with limits from the CMB. The DM self-scattering cross section is proportional to $Q^4$, and we find it to be irrelevant for the phase space. Finally DM-photon scattering is also proportional to $Q^4$ and is not enhanced in the low-velocity limit, so it is also irrelevant.

\subsection{DM with dark photon interactions \label{sec:darkphoton}}

We next consider Dirac fermion DM coupled to a kinetically mixed dark photon $A'$, with the vacuum Lagrangian given by
\begin{align}
	{\cal L} \supset & -\frac{1}{4} F_{\mu \nu} F^{\mu \nu}  +\frac{\kappa}{2} F_{\mu \nu} F^{\prime \mu \nu} -\frac{1}{4} F^\prime_{\mu \nu} F^{\prime \mu \nu} + \frac{1}{2} m_{A'}^2 A_\mu^\prime A^{\prime \mu} \nonumber \\
   &+ \, e J^\mu_{\rm EM} A_\mu + g_\chi\bar \chi \gamma^\mu \chi A^\prime_{\mu} + \bar \chi (i \partial - m_\chi) \chi, \label{eq:vacuumL}  
\end{align}
where $A$ is the SM photon, $\kappa$ is the kinetic mixing parameter and $\chi$ is Dirac fermion DM. For the purposes of this discussion, we consider Abelian kinetic mixing, noting that non-Abelian kinetic mixing is also possible~\cite{Barello:2015bhq,Arguelles:2016ney}. The mixing parameter $\kappa$ could have any number of origins; for instance, it could be generated as a result of loop diagrams with heavy matter fields charged under both $A$ and $A'$ \cite{Dienes:1996zr} or from certain compactifications of type IIB strings \cite{Abel:2008ai,Goodsell:2009xc}. Since the kinetic mixing term is a marginal operator, we take the point of view of a bottom-up effective field theory and we will treat it here as a small free parameter without specifying its origin. 
In this model, the combination of couplings relevant for the relic abundance is $g_\chi g_\text{SM} = g_\chi \kappa e$.

As discussed in Section~\ref{vectors}, the dark photon mass must satisfy $m_{A'} \lesssim 1$~eV in order to give a sufficient coupling for freeze-in while also evading existing bounds on stellar energy loss~\cite{Jaeckel:2013ija}. However, the requirements are even more stringent because unlike the model presented in Section~\ref{sec:millicharge} there could be large $A'$-mediated DM self interaction. For $m_{A'} <$~eV, the mediator would be light enough to give rise to $v^{-4}$ enhanced DM self-scattering in astrophysical environments, with a rate proportional to $g_\chi^4$. Furthermore, as mentioned before, the freeze-in relic abundance is determined by the product $g_\chi \kappa e$, meaning that large $g_\chi$ can be compensated by reducing $\kappa$ to give the same observed relic abundance. Thus a sizable DM self-interaction is possible, and could be relevant to astrophysical probes of self-interacting DM (SIDM). The effects of SIDM are typically parameterized by the momentum-transfer self-scattering cross section, which in the limit of a very light vector mediator is given by~\cite{Feng:2009hw} 
\begin{align}
\sigma_{T,\, \chi \chi} &= \int d\cos \theta_\text{CM} \,  \frac{d\sigma_{\chi \chi}}{d\cos \theta_\text{CM}} (1 - \cos \theta_\text{CM}) \approx \frac{8\pi\alpha_\chi^2}{m_\chi^2 v^4} \ln \frac{(m_\chi v)^2}{m_{A'}^2},
 \label{eq:transfer}
\end{align}
where $\theta_\text{CM}$ is the scattering angle in the center-of-mass (CM) frame, $\sigma_{\chi\chi}$ is the self-interaction cross section, and $\alpha_\chi$ is the dark equivalent of the electromagnetic fine structure constant, $\alpha_\chi \equiv g_\chi^2/4 \pi$.
Typical bounds on SIDM require $\sigma_{\chi \chi}/m_\chi < 1-10$ cm$^2$/g for systems ranging from dwarf galaxies where $v\sim 10^{-4}$ to merging clusters where $v\sim 10^{-2}$ (for a recent review, see Ref.~\cite{Tulin:2017ara}). While few simulation-based studies of self-interactions have been done in the ultralight mediator limit (see for instance Ref.~\cite{Kummer:2019yrb}), we can estimate the expected bound. 
Taking the more restrictive limit of $\sigma_{\chi \chi}/m_\chi\sim$1~cm$^2/$g, the bound is 
\beq g_\chi \lesssim 4 \times 10^{-5}\times \left( \frac{v}{10^{-3}}\right) \times \left(\frac{m_\chi}{1\,\text{MeV}}\right)^{3/4}\times \left(\frac{10}{\ln \left(m_\chi^2 v^2/m_{A'}^2\right) }\right)^{1/4}. \label{SIDMbound}
\eeq
Since $\kappa e g_\chi \gtrsim 10^{-12}$ is needed for sub-MeV freeze-in, the SIDM bounds imply that the kinetic mixing is $\kappa  \gtrsim 10^{-7}$ for MeV-scale DM. For sub-eV dark photons, such large kinetic mixing is only possible when $m_{A'} \lesssim 10^{-10}$ eV~\cite{Jaeckel:2013ija}. For even lighter DM, $g_\chi$ is even more restricted so $\kappa \gtrsim 10^{-5}$ is required for freeze-in, which is possible when $m_{A'}\lesssim 10^{-14}$~eV. Therefore, we are required to consider an ``ultralight'' dark photon~\cite{Knapen:2017xzo}. Note that black hole superradiance constrains dark photons being present in the mass spectrum (in the small-coupling limit) between $\sim 10^{-14} - 10^{-11}$~eV and preliminarily between $\sim 10^{-19} - 10^{-17}$~eV~\cite{Baryakhtar:2017ngi}.

Such a light dark photon is phenomenologically equivalent to the massless dark photon limit for all processes considered in this paper because the $m_{A'}$ is much lower than the effective in-medium photon mass $m_A$ in the early universe. Then, following Appendix D of Ref.~\cite{Knapen:2017xzo}, the vacuum Lagrangian in Eq.~\eqref{eq:vacuumL} is modified with an additional term $m_{A}^2 A^\mu A_\mu /2$.\footnote{For simplicity we consider a constant $m_A^2$ for the schematic purposes of this discussion, although the photon polarization tensor $\Pi^{\mu \nu}(\vec{q}, \omega)$ (which gives rise to the in-medium effective mass) depends on the photon momentum $\vec{q}$, energy $\omega$, polarization, and thermal properties of the medium. For an on-shell mode with $\omega \sim |\vec{q}|$, $m_A^2$ would correspond to the plasma mass, as discussed in Section~\ref{sec:plasmon}. For scattering processes with a highly off-shell mode, $|\vec{q}| \gg \omega$, $m_A^2$ is given by the Debye mass~\cite{Blaizot:1995kg}.} Rotating away the mixing term in the presence of $m_{A}$ and $m_{A'}$ and rewriting in terms of the mass eigenstates $\tilde A$ and $\tilde A'$, the in-medium Lagrangian is given by
\begin{align}
	{\cal L}_{\rm IM} \supset &  -\frac{1}{4} \tilde F_{\mu \nu} \tilde F^{\mu \nu} 
			- \frac{1}{4} \tilde F'_{\mu \nu} \tilde F'^{\mu \nu} 
			+ \frac{m_{A}^2}{2} \tilde A^\mu \tilde A_\mu
			+ \frac{m_{A'}^2}{2} \tilde A'^\mu \tilde {A'}_\mu \nonumber \\
	 &+ J_{\rm EM}^\mu \left(e \tilde A_\mu + \frac{e \kappa \, m_{A'}^2}{m_{A'}^2 - m_{A}^2} \tilde {A'}_\mu\right)    + g_\chi \bar \chi \gamma^\mu \chi \left(  \tilde {A^\prime}_\mu - \frac{\kappa m_{A}^2}{m_{A'}^2 - m_{A}^2}  \tilde A_\mu \right).
	\label{eq:inmedium_L}
\end{align}
From this, we see that when $m_{A} \gg m_{A'}$, the interaction terms above reduce to
\begin{align}
	{\cal L}_{\rm IM} \supset J_{\rm EM}^\mu \left(e \tilde A_\mu \right)    + g_\chi   \bar \chi \gamma^\mu \chi  \left(\tilde {A^\prime}_\mu + \kappa \tilde A_\mu \right),
	\label{eq:inmedium_interactions}
\end{align}
meaning that DM has an effective millicharge parameter $Q = \kappa g_\chi/e$, and the interactions are identical to those for a massless dark photon. Note that this suppression of the $A'$-SM coupling in the $m_{A'} \ll m_A$ limit is the source of the in-medium (plasma mass) suppression of the stellar constraints on dark photons~\cite{An:2013yfc,An:2013yua} discussed in Section~\ref{vectors}. Also note that this suppression means that the dark photon is not abundantly produced by SM interactions in the early universe and does not contribute to the effective number of relativistic species, $N_\text{eff}$.

In the exactly massless $A'$ limit, we are free to perform a field redefinition on $A' \to A' + \kappa A$ in the vacuum Lagrangian, Eq.~\eqref{eq:vacuumL}, which eliminates the kinetic mixing term and generates a DM interaction term $g_\chi \bar \chi \gamma^\mu \chi (A^\prime_\mu + \kappa A_\mu)$, which is again identical to having a millicharge $Q = \kappa g_\chi/e$ under $U(1)_{EM}$. 

The model considered here thus provides another realization of millicharged DM, and all of the stellar constraints discussed in the previous section apply. The only difference is the additional DM self-interaction via the $A'$, which potentially leads to sizeable self-interactions.

\section{Relic abundance from freeze-in \label{sec:prod}}

Here we compute the relic abundance of DM from freeze-in. We begin by reproducing the contribution from annihilation of SM fermions $f \bar f \to \chi \bar \chi$ that was previously calculated in Refs.~\cite{Essig:2011nj,Chu:2011be}. Because freeze-in is peaked at low temperatures and this paper concerns sub-MeV DM, electrons are the primary source of DM for this channel; in the rest of this section we explicitly refer to freeze-in off electrons, noting we have numerically checked that adding heavier fermions (for instance muons) to the calculation changes the results by less than 1\%. In addition to freeze-in off electrons, there is a contribution from plasmon decays, $\gamma^* \to \chi \bar \chi$, which we calculate for the first time. Photon annihilation into DM $\gamma \gamma \to \chi \bar \chi$ is suppressed by an additional factor of $Q^2$ and can be safely neglected. 

In what follows, we take the observed present-day relic DM abundance to be $\omega_{c} \equiv \Omega_{c} h^2 = 0.12$~\cite{Aghanim:2018eyx}. After freeze-in, the DM density should scale like $a^{-3}$ and it is common practice to compare this to another quantity that has the same scaling irrespective of changes to the SM bath temperature. In this work we choose to compare the number density to the entropy density. Taking the present-day CMB temperature to be 2.73~K, the observed yield is then \beq Y\equiv n_\chi/s = 4.35\times 10^{-7} \times \left(\frac{1\, \mathrm{ MeV}}{m_\chi}\right) .\eeq
For $m_\chi\gtrsim1$~keV, the DM yield is much lower than the order unity yield for relativistic species, such that DM contributes negligibly to $N_\text{eff}$. This is in contrast to other DM models, such as thermal freeze-out, where sub-MeV DM would generically inject a considerable amount of entropy to the photon or neutrino sectors and would violate observational bounds on $N_\text{eff}$.\footnote{An exception for thermal, sub-MeV DM was pointed out in Ref.~\cite{Berlin:2017ftj}, where the DM thermalizes with the SM thermal bath {\emph{after}} neutrino-photon decoupling, reducing the contribution to $N_\text{eff}$. Furthermore, in this model changes to $N_\text{eff}$ that occur after DM thermalization are compensated by decoupling at a later time.}

The low DM occupation number also implies that it is possible to self-consistently ignore back-reactions that would reduce the DM number density, namely DM annihilation to electrons and inverse decays to plasmons. For instance, if we ignore the back-reaction, the solution for the number density of DM is significantly lower than the electron number density during the entirety of freeze-in in spite of the fact that the latter is becoming Boltzmann suppressed. Depletion of the DM number density through annihilation to dark photons $\chi \bar \chi \to \gamma^{'} \gamma^{'}$ is negligible for the same reason. In what follows, we solve the $0^\text{th}$ moment of the Boltzmann equation ignoring back-reactions, noting that we have numerically checked that they are negligible. The relevant equation is then
 \beq 
   \frac{d n_\text{DM}}{d a} + \frac{3 n_\text{DM}}{a} = \frac{2}{a H}\left( \avg{\sigma v}_{e^+ e^- \rightarrow \chi\bar{\chi}} n_e^2
  + \avg{\Gamma}_{\gamma^{*} \rightarrow \chi \bar{\chi}} n_{\gamma^*}\right).
  \label{0thboltz}
  \eeq
Here we are using $a$ as our time variable. The relationship between $a$ and the SM temperature $T$ (which determines the DM production rate) is not adiabatic during freeze-in because the electrons are leaving the thermal bath at this time; this is discussed further in Appendix~\ref{clock}. Note that we are solving for the total DM density which includes both $\chi$ and $\bar{\chi}$ in the matter budget; assuming zero DM chemical potential, $n_\text{DM} = 2 n_\chi = 2 n_{\bar{\chi}}$, which accounts for the factor of two in Eq.~\eqref{0thboltz}.\footnote{This factor is related to the usual factor of $1/2$ that appears in the Boltzmann equation for Dirac fermions~\cite{Gondolo:1990dk, Srednicki:1988ce}; however, unlike the ordinary case of thermal DM, the change in the comoving DM density for freeze-in is independent of the DM number density (i.e. there is no factor of $n_\text{DM}^2$ appearing in Eq.~\eqref{0thboltz}) which accounts for the factor of four difference.}

\subsection{Annihilations}

In computing the DM relic abundance from annhilations of electron-positron pairs, we treat the two scenarios discussed in Section~\ref{sec:model} as indistinguishable in the limit that $m_{A'}\rightarrow 0$. We also ignore the in-medium photon mass for this process, which we find to be a percent level effect for $s$-channel annihilations happening at the relevant range of temperatures. In this limit, the matrix element squared is
 \begin{align*} 
    \sum_\mathrm{d.o.f.}\abs{\mathcal{M}}^2_{e^+ e^- \leftrightarrow \chi \bar{\chi}} =\frac{32 
 Q^2 e^4}{ (p_{e^+}+p_{e^-})^4 
} \Big(& (p_{e^+}\cdot p_\chi) (p_{e^-}\cdot p_{\bar{\chi}}) + (p_{e^+}\cdot p_{\bar{\chi}}) (p_{e^-}\cdot p_{\chi})\\ &
 + m_e^2 (p_\chi\cdot p_{\bar{\chi}}) + m_\chi^2(p_{e^+}\cdot p_{e^-}) + 2 m_e^2 m_\chi^2 \Big), \numberthis
 \end{align*} 
 where we sum over both initial \emph{and} final spin degrees of freedom (d.o.f.) without averaging and where $Q$ is the effective millicharge in the dark photon case, $Q = \kappa g_\chi /e$.
The thermally averaged cross section appearing in Eq.~\eqref{0thboltz} for this process is given by
\begin{align*} \avg{\sigma v}_{e^+ e^- \rightarrow \chi\bar{\chi}} n_e^2  =  \int \frac{ \dbar^3 p_{e^+}}{2 E_{e^+}} \frac{ \dbar^3 p_{e^-}}{2 E_{e^-}}&  \frac{ \dbar^3 p_\chi}{2 E_\chi}  \frac{ \dbar^3 p_{\bar{\chi}}}{2 E_{\bar{\chi}}} ~e^{-(E_{e^+} + E_{e^-})/T}\numberthis \label{sigmavann}\\& \times \sum_\mathrm{d.o.f.} \abs{\mathcal{M}}_{e^+ e^- \rightarrow \chi\bar{\chi}}^2 (2 \pi)^4 \delta^{(4)}(p_{e^+} + p_{e^-} - p_\chi - p_{\bar{\chi}}) \quad \quad  \end{align*}
where $\dbar^3 p \equiv d^3 p/(2\pi)^3$. 
We assume that from the onset of freeze-in, the electrons have entered the non-relativistic regime where their phase space is given by a Maxwell-Boltzmann distribution with temperature $T$ and zero chemical potential. As we will show, sub-MeV DM freeze-in through the annihilation channel is most effective at temperatures $T\lesssim m_e$ where the effects of Fermi-Dirac statistics can be neglected. We also ignore Pauli blocking of the DM due to its low occupation number.

 To evaluate the thermal cross section, we note that the primordial plasma has a preferred rest frame (where bulk motions average to zero), which breaks Lorentz invariance. The phase space factors of Eq.~\eqref{sigmavann} are evaluated in a frame that is comoving with the plasma. Practically, we can perform the integration by inserting factors of unity, 
\beq 
    \int \frac{ d^3 q_{12} d s_{12} }{2 E_{12}} \delta^{(4)}(q_{12} - p_1 - p_2)=1, 
\label{eq:unity}
\eeq 
 where $q_{12}$ is the effective bulk 4-momentum of the particles labelled 1 and 2 and $s_{12}$ can be thought of as the effective (Lorentz invariant) mass-squared of a single particle with that bulk 3-momentum and energy (\emph{i.e.} here $E_{12} = \sqrt{s_{12} + \vec{q}_{12}^{\,2}}$). Inserting such a factor into Eq.~\eqref{sigmavann} gives
 \begin{align*} \avg{\sigma v}_{e^+ e^- \rightarrow \chi\bar{\chi}} n_e^2  &= 
 \int \frac{ d^3 q_{\chi \bar{\chi}} d s_{\chi \bar{\chi}} }{2 E_{\chi \bar{\chi}}} \int \frac{ \dbar^3 p_{e^+}}{2 E_{e^+}} \frac{ \dbar^3 p_{e^-}}{2 E_{e^-}} \frac{ \dbar^3 p_\chi}{2 E_\chi}  \frac{ \dbar^3 p_{\bar{\chi}}}{2 E_{\bar{\chi}}} ~e^{-(E_{e^+} + E_{e^-})/T}\numberthis \\&  \times \sum_\mathrm{d.o.f.} \abs{\mathcal{M}}_{e^+ e^- \rightarrow \chi\bar{\chi}}^2 (2 \pi)^4 \delta^{(4)}(p_{e^+} + p_{e^-} - p_\chi - p_{\bar{\chi}}) \delta^{(4)}(q_{\chi \bar{\chi}} - p_\chi - p_{\bar{\chi}}).\quad \quad  \end{align*}
The integral over $p_\chi$ and $p_{\bar{\chi}}$ does not depend on the frame of $q_{\chi \bar{\chi}}$, so the two-body phase space of $p_\chi$ and $p_{\bar{\chi}}$ can be evaluated in the CM frame of $q_{\chi \bar{\chi}}$. We define
 \begin{align*} &\Phi_{\chi \bar{\chi}}(s_{\chi \bar{\chi}}) \abs{\mathcal{M}}^2_\text{CM}(s_{\chi \bar{\chi}}) \equiv \int \frac{ \dbar^3 p_\chi}{2 E_\chi}  \int \frac{ \dbar^3 p_{\bar{\chi}}}{2 E_{\bar{\chi}}} (2 \pi)^4 \delta^{(4)}(q_{\chi \bar{\chi}} - p_\chi - p_{\bar{\chi}}) \sum_\mathrm{d.o.f.} \abs{\mathcal{M}}_{e^+ e^- \rightarrow \chi\bar{\chi}}^2 \\
& = \frac{ Q^2 e^4}{2 \pi s_{\chi \bar{\chi}}^2 } \sqrt{1 -\frac{4 m_\chi^2}{s_{\chi \bar{\chi}}}}  \left(  s_{\chi \bar{\chi}}^2 + \frac{1}{3}(s_{\chi \bar{\chi}} - 4 m_e^2) (s_{\chi \bar{\chi}} - 4 m_\chi^2)+  4 s_{\chi \bar{\chi}} (m_\chi^2 +m_e^2) \right),\quad \quad \quad \numberthis \end{align*}
 and insert this into the expression for the thermally averaged cross section
 \begin{align*} \avg{\sigma v}_{e^+ e^- \rightarrow \chi\bar{\chi}}  n_e^2  =  \int \frac{ d^3 q_{\chi \bar{\chi}} d s_{\chi \bar{\chi}} }{2 E_{\chi \bar{\chi}}}~ e^{-E_{\chi \bar{\chi}}/T} & \Phi_{\chi \bar{\chi}}(s_{\chi \bar{\chi}})\abs{\mathcal{M}}^2_\text{CM}(s_{\chi \bar{\chi}})\\
 \times & \int \frac{ \dbar^3 p_{e^+}}{2 E_{e^+}} \frac{ \dbar^3 p_{e^-}}{2 E_{e^-}} \delta^{(4)}(p_{e^+}+p_{e^-}-q_{\chi \bar{\chi}} ) . \numberthis \end{align*}
Again, we can evaluate the integral over $p_{e^+}$ and $p_{e^-}$ in the center-of-mass frame. Defining
\beq \Phi_{e^+ e^- }(s_{\chi \bar{\chi}}) \equiv \frac{1}{8 \pi} \sqrt{1 - \frac{4 m_e^2}{s_{\chi \bar{\chi}}}},\eeq
the thermally averaged cross section becomes
  \begin{align}
\avg{\sigma v}_{e^+ e^- \rightarrow \chi\bar{\chi}}  n_e^2 &= \frac{1 }{ (2 \pi)^4}  \int \frac{ d^3 q_{\chi \bar{\chi}} d s_{\chi \bar{\chi}} }{2 E_{\chi \bar{\chi}}}  e^{-E_{\chi \bar{\chi}}/T}  \Phi_{e^+ e^- }(s_{\chi \bar{\chi}})\Phi_{\chi \bar{\chi}}(s_{\chi \bar{\chi}})\abs{\mathcal{M}}^2_\text{CM}(s_{\chi \bar{\chi}}).
\end{align}
We can write this result in terms of the first order modified Bessel function of the second kind $K_1(z) = z \int_1^\infty d u \,e^{- z u} \sqrt{u^2 - 1} $ with $u = \sqrt{1+ q_{\chi \bar{\chi}}^2 /s_{\chi \bar{\chi}}} \,$ :
\beq \avg{\sigma v}_{e^+ e^- \rightarrow \chi\bar{\chi}}  n_e^2 = \frac{ T}{(2 \pi)^3} \int ds\,\sqrt{s}~ \Phi_{e^+ e^- }(s)\,\Phi_{\chi \bar{\chi}}(s) \abs{\mathcal{M}}^2_\text{CM}(s)\, K_1(\sqrt{s}/T)\eeq
where we have dropped the subscript on the integration variable $s$. Note that $s$ is restricted to $s > 4\max\left(m_e^2, m_\chi^2\right)$. The procedure above provides an alternate derivation of the well-known results from Ref.~\cite{Gondolo:1990dk}, and we have validated this method here because we use similar techniques to derive the full collision term for annihilation in Section~\ref{sec:ann_phasespace}.

\subsection{Plasmon decay \label{sec:plasmon}}

The early Universe is an optically thick plasma where photons acquire an in-medium mass; this can be understood classically as arising from the electrons' oscillatory response to a propagating electric field and the dynamical shielding of that electric field. This effective mass is also manifest in the photon propagator and the polarization vectors of external photon legs in the medium; in other words, the photon mass and wavefunction are renormalized in the plasma. The effective masses and dressed polarization functions for the transverse and longitudinal ``plasmon'' modes are shown in Fig.~\ref{fig:plasmon} and explicit formulae are provided in Appendix~\ref{plasma}.  The effective mass for plasmons is closely related to the plasma frequency. For a relativistic plasma at zero chemical potential, the plasma frequency is $\omega_p = eT/3 \approx 0.1 T$ where $e$ is electric charge. 

Plasmons can undergo decay provided that it is kinematically allowed. For instance, plasmons can decay to neutrino pairs through mixing with the $Z$ boson \cite{Braaten:1993jw}. Plasmons cannot decay to charged particles in the SM because their effective mass is also renormalized in the medium and it is always kinematically forbidden. However, this is not the case for millicharged DM where corrections to the mass are suppressed by powers of $Q$.  

\onecolumngrid
\begin{figure*}[t!]
\includegraphics[width=\textwidth]{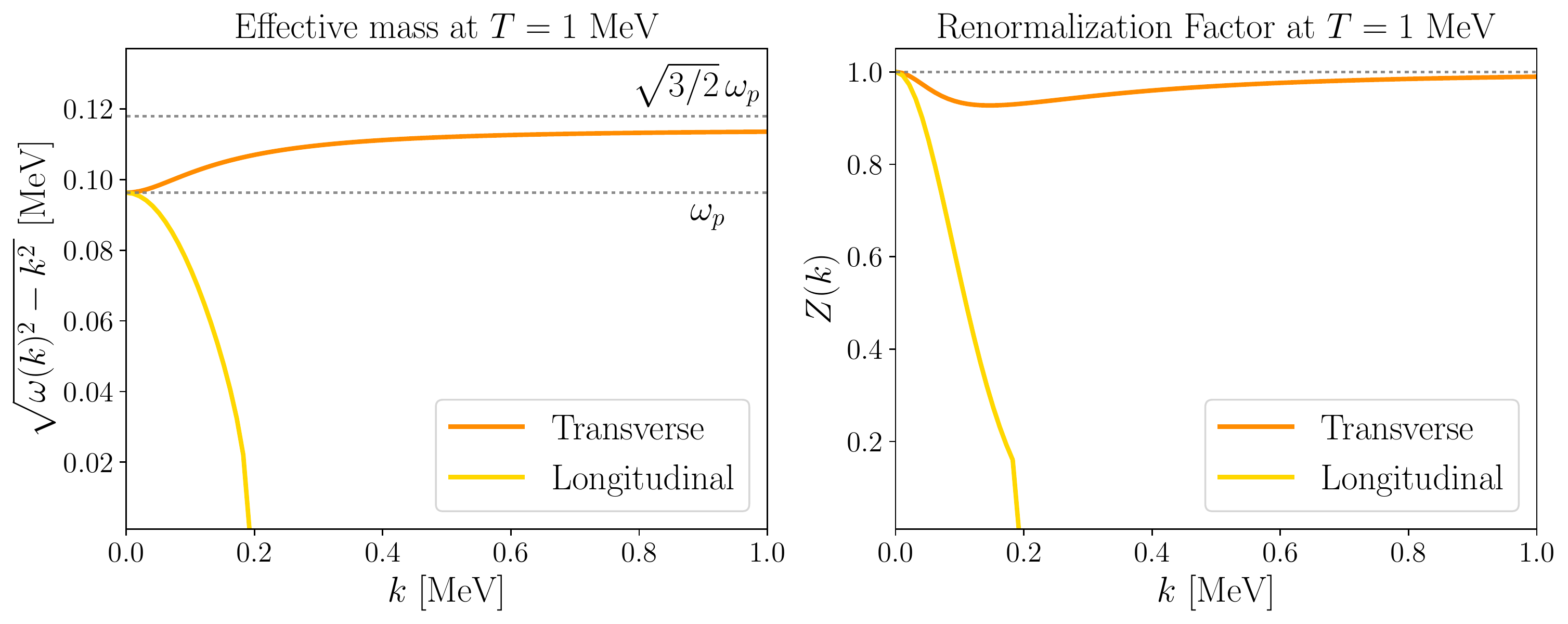}
\caption{The effective in-medium mass (left) and wavefunction renormalization (right) for photons, as computed in Coulomb gauge for a plasma with $T=1$~MeV and zero chemical potential (see Appendix~\ref{plasma} for relevent formulae). The transverse mode is relevant at all wavelengths while the longitudinal mode crosses the lightcone at high $k$ and can thus only propagate at low $k$. Also shown are the low-$k$, low-$T$ and high-$k$, high-$T$ limits for the effective transverse mass, $m_t = \omega_p$ and $m_t = \sqrt{3/2}\omega_p$, respectively.}
\label{fig:plasmon}
\end{figure*}

The effective matrix element that captures plasmons decaying to DM is 
\beq  
i \mathcal{M}_{\gamma^* \rightarrow \chi \bar{\chi}} = i Q e\, \tilde{\epsilon}_\mu(k) \bar{u}(p_\chi) \gamma^\mu v(p_{\bar{\chi}}), \eeq 
where $\tilde{\epsilon}_\mu(k)$ is the dressed polarization vector for the longitudinal and transverse plasmon modes as detailed in Appendix~\ref{plasma}, where we work in Coulomb gauge. We express this process in terms of the DM effective millicharge $Q$ and in Appendix~\ref{basis} we show explicitly that decaying through a dark photon gives the same effective matrix element in the limit $m_{A'}\rightarrow 0$. In squaring and summing over polarizations, only the diagonal terms ($LL$, $++$, and $--$) contribute,
\beq \sum_\mathrm{d.o.f.} \abs{\mathcal{M}}^2_{\gamma^* \rightarrow \chi \bar{\chi}} = 4 Q^2 e^2 \times \begin{cases}
2 Z_t(k) ( p_\chi^2 \sin^2 \theta + \omega_t(k) E_\chi - k p_\chi \cos \theta) & {++ \& --} \\
Z_\ell(k) \frac{\omega_\ell(k)^2}{k^2} (\omega_\ell(k) E_\chi - 2 E_\chi^2 + k p_\chi \cos \theta) & \text{LL},
\end{cases}
\eeq
where the photon four-momentum is given by $K^\mu = \big(\omega(k), \vec{k}\big)^\mu$ with appropriate dispersion relations for transverse and longitudinal modes $\omega_t(k)$ and $\omega_\ell(k)$ (see Appendix~\ref{plasma}), the DM four-momentum is given by $\left(E_\chi, \vec{p}_\chi\right)^\mu$, $\theta$ is the angle between $\vec{k}$ and $\vec{p}_\chi$, and $Z_t(k)$ and $Z_\ell(k)$ are wavefunction renormalization factors (shown in Fig.~\ref{fig:plasmon}) that are related to the dressed polarization vectors for the transverse and longitudinal modes.

The thermally averaged decay rate is \beq \left< \Gamma\right>_{\gamma^* \rightarrow \chi \bar{\chi}} n_{\gamma^*} = \int \frac{\dbar^3 k}{2 \omega(k)}  \frac{\dbar^3 p_\chi}{2 E_\chi}\frac{\dbar^3 p_{\bar{\chi}}}{2 E_{\bar{\chi}}} ~f\left(\omega(k)\right) 
 (2\pi)^4 \delta^{(4)}\left(K - p_\chi - p_{\bar{\chi}}\right) \sum_\mathrm{d.o.f.} \abs{\mathcal{M}}^2_{\gamma^* \rightarrow \chi \bar{\chi}},\eeq
and can be evaluated directly. Taking the plasmons to be Bose-Einstein distributed, the longitudinal and transverse contributions to this rate are
\beq \left< \Gamma\right>_{\gamma^*_\ell \rightarrow \chi \bar{\chi}} n_{\gamma^*_\ell} =\frac{ Q^2 e^2}{(2 \pi)^3} \int k^2 \,d k  \, \frac{Z_\ell(k) \omega_\ell (k)(m_\ell(k)^2 + 2 m_\chi^2 )\sqrt{m_\ell(k)^2(m_\ell(k)^2 -4 m_\chi^2)}}{3 m_\ell(k)^4 \left(e^{\omega_\ell(k)/T}-1 \right)}  \eeq
\beq \left< \Gamma\right>_{\gamma^*_t \rightarrow \chi \bar{\chi}} n_{\gamma^*_t}= \frac{ 4Q^2 e^2 }{(2 \pi)^3} \int k^2 \,d k  \, \frac{ Z_t(k)(m_t(k)^2 -m_\chi^2)\sqrt{m_t(k)^2 (m_t(k)^2 - 4 m_\chi^2)}}{3 \omega_t (k)\,m_t(k)^2 \left(e^{\omega_t(k)/T}-1 \right)},  \eeq
where the effective plasmon masses are $m_\ell(k)^2 = \omega_\ell(k)^2 - k^2$ for the longitudinal modes and $m_t(k)^2 = \omega_t(k)^2 - k^2$ for the tranverse ones.
The final integrals over $k$ can be computed numerically and the total plasmon contribution to decay is dominated by the transverse modes (note that we are working in Coulomb gauge). This is because the longitudinal mode has a finite range of $k$ over which it can propagate, meaning that it has less available phase space than the transverse mode which has no restriction in $k$. Furthermore, the longitudinal mass and renormalization factors fall steeply within the range of $k$ where this mode can propagate.

\subsection{Couplings for freeze-in}

\begin{figure*}[t]
\includegraphics[width=0.5\textwidth]{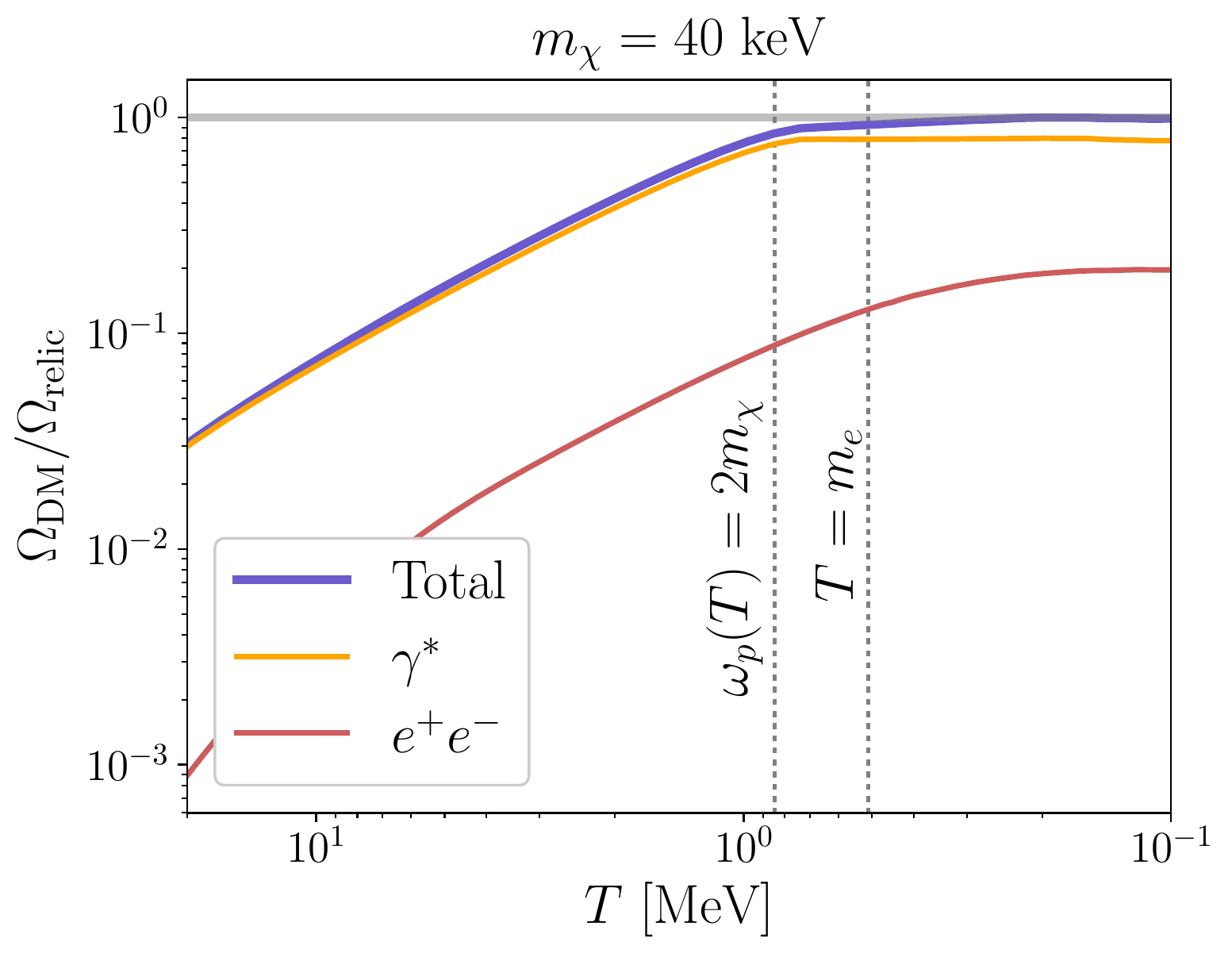}\includegraphics[width=0.5\textwidth]{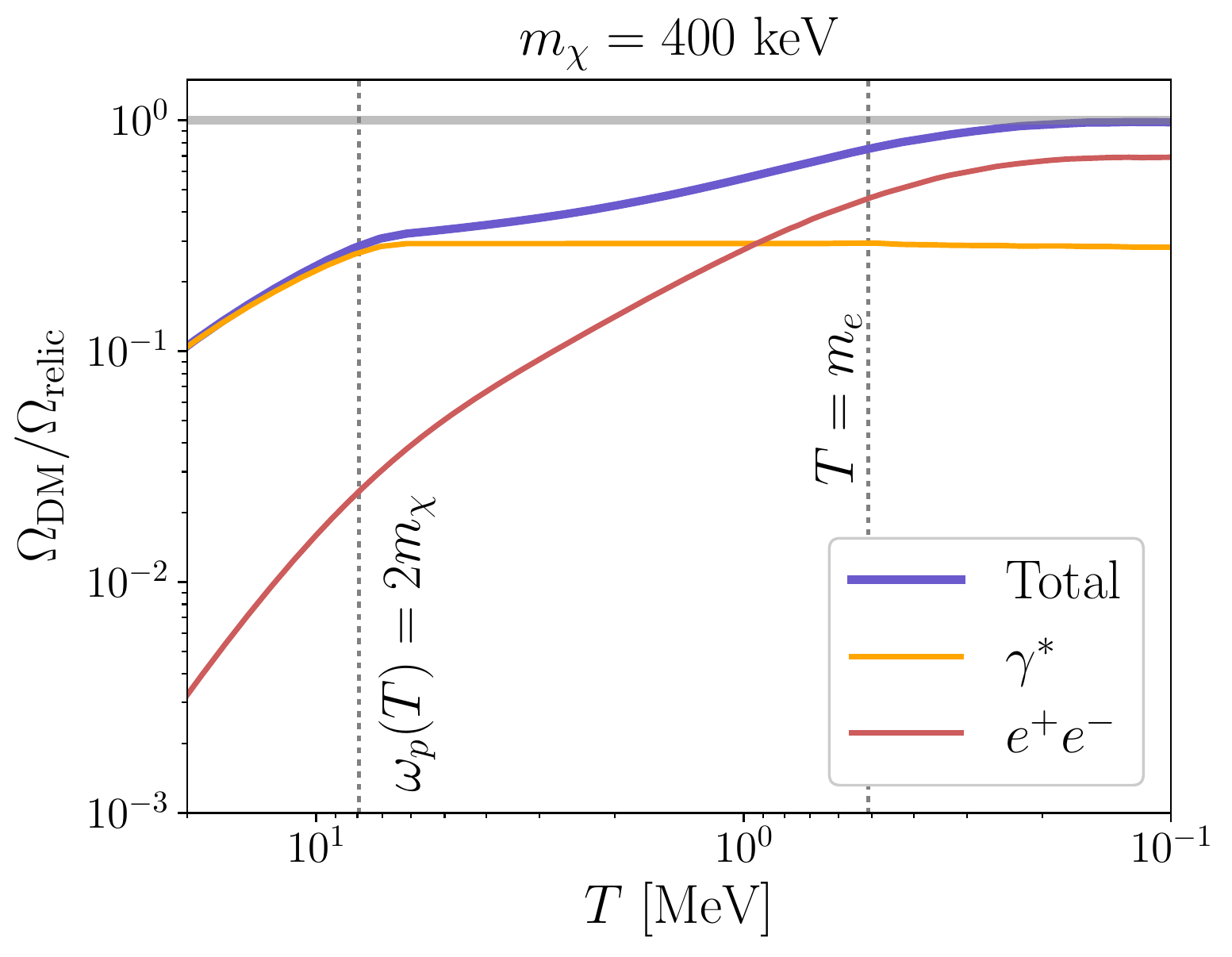}
\caption{Evolution of the comoving DM number density for $m_\chi = 40$~keV (left) and $m_\chi = 400$~keV (right) as compared to the relic abundance of DM with that mass. Also shown are the relative contributions from electron-positron annihilations and plasmon decays, as discussed in the text.}
\label{masscomp_abundance}
\end{figure*}

In solving the zeroth moment of the Boltzmann equation for the DM relic abundance, we find that the relative contributions from $e^+ e^-$ annihilation and plasmon decays are starkly different in different mass ranges, as illustrated in Fig.~\ref{masscomp_abundance}. 
This can be understood by considering the fact that freeze-in is dominant at low temperatures, provided that it is kinematically allowed and that the population the DM is freezing in from has a sufficient abundance. For sub-MeV DM, freeze-in from $e^+e^-$ annihilation is always kinematically allowed and this process only ends when the electron number density becomes Boltzmann suppressed, namely $T\lesssim m_e$. Meanwhile, the plasmon abundance is not Boltzmann suppressed but the mass runs with temperature, so freeze-in through plasmon decay ends when it is no longer kinematically allowed, namely when $m_{\gamma^*}\sim \omega_p = 2 m_\chi$. Since $\omega_p \approx 0.1 T$ in the relativistic limit, plasmon decay to millicharged DM shuts off at an earlier time compared to annihiliation. These two criteria are shown in Fig.~\ref{masscomp_abundance} and indeed we see that plasmon decays are more dominant in determining the relic abundance for lower mass DM because the decays are active for a longer period of time.

\begin{figure}[t]
\includegraphics[width=0.7\textwidth]{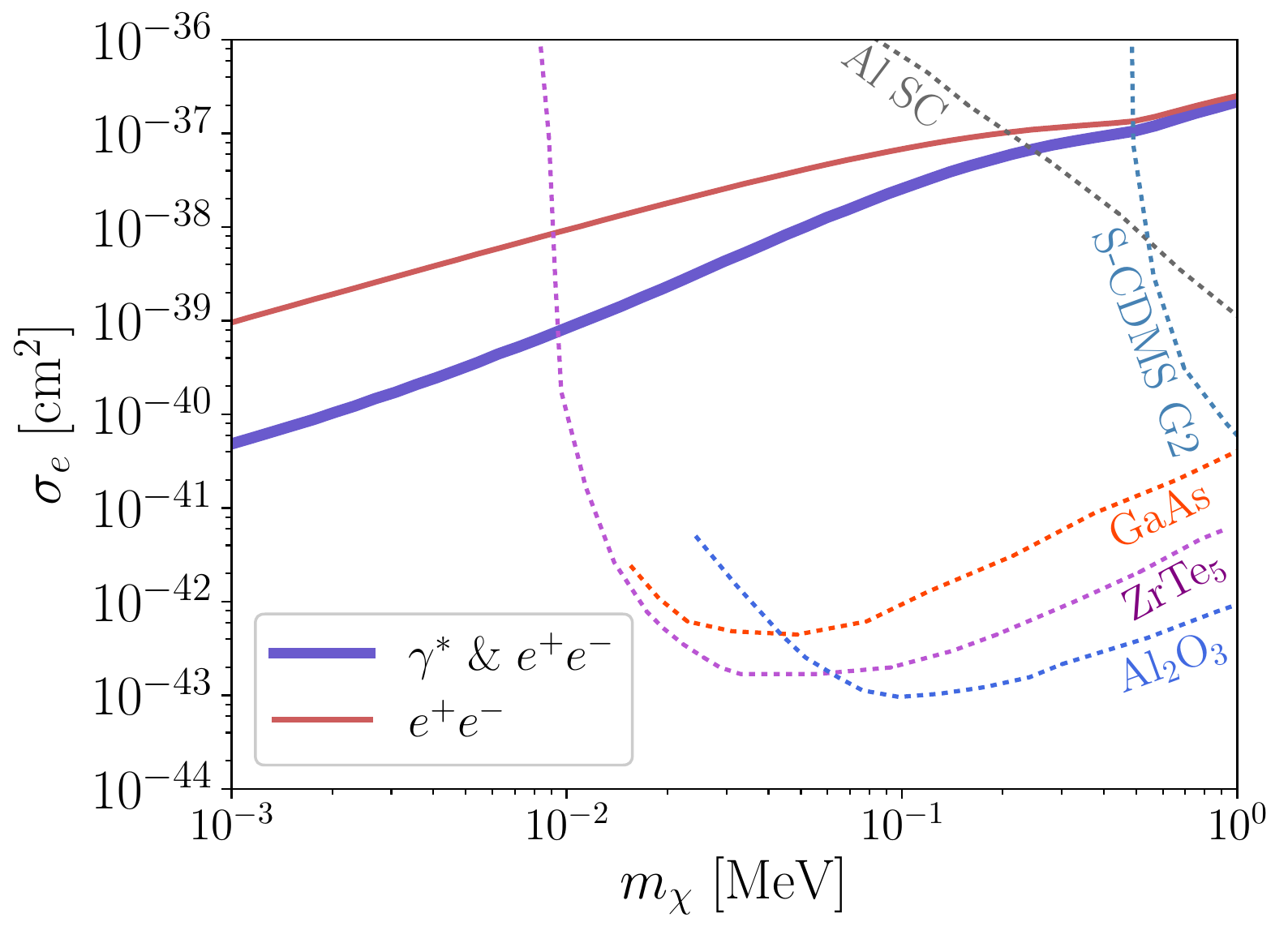} 
\caption{The effect of plasmon decays on the freeze-in benchmark for direct detection via electron recoils. Also shown are the projected sensitivities of low-threshold experiments with kg-day exposure, including a SuperCDMS G2 experiment~\cite{Battaglieri:2017aum} and proposals using polar materials (GaAs and Al$_2$O$_3$)~\cite{Griffin:2018bjn, Knapen:2017ekk}, Dirac materials (ZrTe$_5$)~\cite{Hochberg:2017wce}, or superconductors (Al SC)~\cite{Hochberg:2015fth}.
\label{DDshift}}
\end{figure}

In terms of the effective millicharge needed to produce the observed DM relic abundance, we find that including plasmon decays leads to a significant reduction in coupling for keV-mass DM while the effect is small once $m_\chi$~=~MeV. 
The change to the freeze-in benchmark for direct detection is shown in Fig.~\ref{DDshift},
where the cross section for electron recoils is 
\beq \sigma_e = \frac{16 \pi Q^2 \alpha^2 \mu_{\chi e}^2}{ (\alpha m_e)^4} . \eeq
Here $\mu_{\chi e}$ is the electron-DM reduced mass, $\mu_{\chi e} = m_e m_\chi / (m_e + m_\chi)$. At the lowest mass where proposed low-threshold direct detection experiments are sensitive, the plasmon decay channel for DM production lowers the expected signal strength by roughly an order of magnitude.

It has been noted in the literature~\cite{Chuzhoy:2008zy,Hu:2016xas,Dunsky:2018mqs} that millicharged DM could be efficiently accelerated in supernova remnants, which would lead to an accelerated component of dark cosmic rays and  eject DM from the disk. Both of these effects can lead to substantial changes to the predicted direct detection rates and sensitivities of proposed experiments shown above. However, the conclusions are highly sensitive to aspects of cosmic ray physics which are not fully understood, such as the injection of particles into the diffusive shock acceleration process. The predictions would also be sensitive to whether the DM obtains its effective millicharge through a kinetic mixing portal; in this case, the dark photon mass and couplings can affect the acceleration, and an exploration of these effects is beyond the scope of this work.

\section{Dark matter phase space distribution \label{sec:phasespace}}

Since freeze-in DM is so weakly coupled to the SM, it does not thermalize with the SM during freeze-in and the phase space distribution can deviate substantially from a thermal distribution. While this has no clear impact on direct detection, since galaxy assembly is expected to significantly alter the DM velocity distribution, it does affect DM free-streaming and DM-SM scattering in the early universe. Here we compute the full phase space distributions needed to determine the cosmological observables; the signatures, constraints, and detection prospects will be presented in a companion paper~\cite{inprep}.

We must solve the full Boltzmann equation in an expanding background, given by  
\beq 
    \frac{\partial f_\chi}{\partial t} - H \frac{p_\chi^2}{E_\chi} \frac{\partial f_\chi}{\partial E_\chi} = \frac{C(p_\chi, t)}{E_\chi}, 
\eeq
where $C(p_\chi, t)$ is the collision term, which encapsulates all interactions that affect the phase space. At early times, the interactions that determine the phase space evolution are $e^+ e^-$ annihilation and plasmon decay. We have checked numerically that heavier fermion annihilation processes (for instance the annihilation of muon-antimuon pairs) affect the phase space by a negligible amount because they occur only at early times when freeze-in is less efficient. Scattering has a negligible impact on the phase space during freeze-in since the DM occupation number is much smaller than that of electrons or plasmons. Neglecting the small effect of scattering during freeze-in, the collision term is independent of $f_\chi$ to leading order and the Boltzmann equation can be solved by direct integration~\cite{Bae:2017dpt}, 
\beq 
    f_\chi(p_\chi, t) = \int_{t_i}^{t} dt'\, \frac{C\left(\frac{a(t)}{a(t')}\, p_\chi, t'\right)}{\sqrt{\frac{a(t)^2}{a(t')^2}\, p_\chi^2+m_\chi^2} } = \int_{a_i}^{a(t)} \frac{da'}{a' H(a')}\, \frac{C\left(\frac{a(t)}{a'}\, p_\chi, a'\right)}{\sqrt{\frac{a(t)^2}{a'^2}\, p_\chi^2+m_\chi^2} }. \label{axinoapprox}
\eeq
Here the factors of $a$ in the integrand keep track of redshifting of momentum due to expansion. We use the scale factor $a$ as our time variable rather than the common choice of using the SM temperature because it is not evolving adiabatically as the electron-positron pairs leave the bath during freeze-in. The temperature evolution and the evolution of the Hubble parameter are detailed in Appendix~\ref{clock}. 

After freeze-in ends, the DM momenta redshift and the phase space distribution is constant in comoving momentum. However, at late times DM-SM and DM-DM scattering eventually can become important since the scattering cross sections are peaked at low relative velocities. The effects of DM-SM scattering on the phase space are generally negligible for the allowed parameter space, but DM self-scattering can lead to thermalization of the DM phase space distribution. Whether this occurs is model-dependent, and we discuss the conditions for this to occur in Section~\ref{sec:DMscattering}.
 
\subsection{Phase space from annihilation \label{sec:ann_phasespace} }
\begin{figure*}[t]
\includegraphics[width=\textwidth]{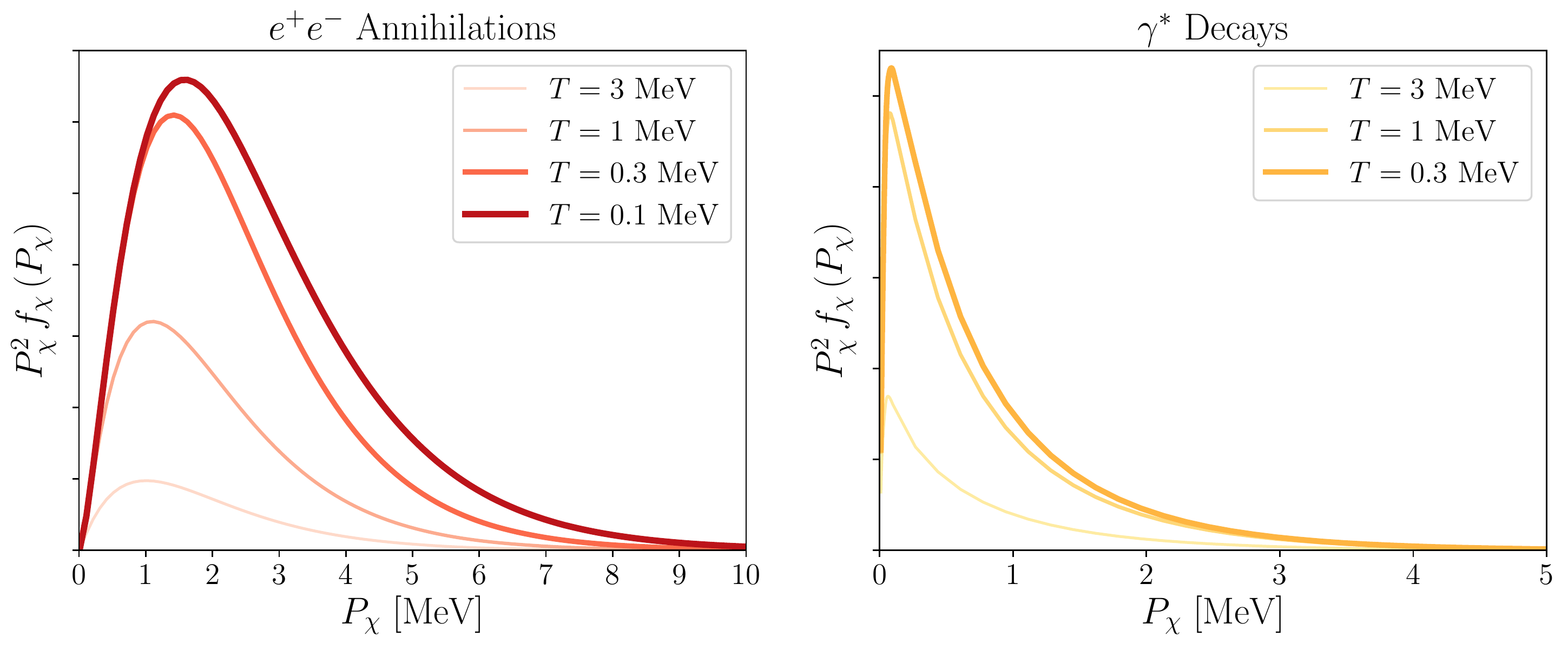}
\caption{A comparison of the phase space evolution of DM being produced by $e^+ e^-$ annihilation (left) and $\gamma^*$ decay (right) at $m_\chi = 40$~keV. The momenta shown here are comoving, $P_\chi \equiv a p_\chi$ where $a=1$ corresponds to $T= 1$~MeV. The phase space is normalized arbitrarily for the purposes of comparing the $P_\chi$-dependence side by side. Over time, the comoving phase space converges to its final frozen-in shape. The phase space from annihilation is similar to that of the thermal electrons from which they inherit their kinematics. Meanwhile, the phase space from plasmon decay is highly peaked at low $P_\chi$ because freeze-in through this channel occurs predominantly at threshold when $\omega_p \sim 2 m_\chi$ and the decay is peaked when the plasmon is ``at rest,'' $k\rightarrow0$. }
\label{freezephase}
\end{figure*}

The computation of the full collision term from annihilation proceeds similarly to the computation of its zeroth moment. Once again, inserting a factor of unity as defined in Eq.~\eqref{eq:unity}, we find 
\begin{align*} 
    C(p_\chi, t)_{e^+ e^- \rightarrow \chi\bar{\chi}} 
    = \frac{1}{2(2 \pi)^3}  \int \frac{ d^3 q_{e^+ e^-} d s_{e^+ e^- } }{2 E_{\bar{\chi}} 2 E_{e^+ e^-}} & \delta(E_{e^+ e^-} - E_\chi - E_{\bar{\chi}}) ~e^{-E_{e^+ e^-}/T}\\
    &\times \Phi_{e^+ e^-}(s_{e^+ e^-}) \abs{\mathcal{M}}^2_\text{CM}(s_{e^+ e^-}), \numberthis
\end{align*}
where $E_{\bar{\chi}} = \sqrt{m_\chi^2 + p_{{\chi}}^2 + q_{e^+ e^-}^2 - 2 p_\chi q_{e^+ e^-} \cos \theta}$, $E_{e^+ e^-} = \sqrt{s_{e^+e^-}+q^2_{e^+ e^-}}$ and $\theta$ is the angle that $\vec{q}_{e^+ e^-}$ makes with the unconstrained, unintegrated $\vec{p}_\chi$. 
Defining $x \equiv \cos\theta$ and dropping the subscript on the bulk electron momentum, we find 
\begin{align}
& C(p_\chi, t)_{e^+ e^- \rightarrow \chi\bar{\chi}}  = \frac{1}{2 (2 \pi)^2 p_\chi} \int \frac{ dx \,q d q\, d s }{ 4 E}  \delta \left(x-\frac{2 E_\chi E -s}{2 p_\chi q}\right) e^{-E/T} \Phi_{e^+ e^-}(s) \abs{\mathcal{M}}^2_\text{CM}(s).
\end{align}
Requiring that $x \in [-1, 1]$ and switching integration variables,
\begin{align*} &C(p_\chi, t)_{e^+ e^- \rightarrow \chi\bar{\chi}}  = \frac{1}{8 p_\chi (2 \pi)^2} \int
d s\int_{\frac{E_\chi s - p_\chi \sqrt{s(s-4 m_\chi^2)}}{2 m_\chi^2}}^{\frac{E_\chi s + p_\chi \sqrt{s(s-4 m_\chi^2)}}{2 m_\chi^2}}  d E \,  e^{-E/T} \Phi_{e^+ e^-}(s) \abs{\mathcal{M}}^2_\text{CM}(s) \\
&= \frac{ T }{ 4 p_\chi (2 \pi)^2}  \int
d s \,  e^{-\frac{ E_\chi s}{2 m_\chi^2 T}} \sinh \left( \frac{ p_\chi \sqrt{s(s-4 m_\chi^2)}}{2 m_\chi^2 T}\right) \Phi_{e^+ e^-}(s) \abs{\mathcal{M}}^2_\text{CM}(s). \numberthis \label{colltermann} \end{align*}
Then, to solve for the final phase space from annihilation, we can combine Eqs.~\eqref{axinoapprox} and \eqref{colltermann}. Note that because $p_\chi$ is fixed (rather than an integration variable), $s$ in the above integral is restricted to $s > \max \left(4 m_e^2, 2 m_\chi (E_\chi + m_\chi)\right)$ unlike in the integral for determining the thermally averaged cross section. The resulting evolution of the phase space distribution is shown in the left panel of Fig.~\ref{freezephase}.

\onecolumngrid

\subsection{Phase space from plasmon decay \label{sec:plasmon_phasespace} }

The collision term from plasmon decay,
\beq C(p_\chi, t)_{\gamma^* \rightarrow \chi \bar{\chi}} =   \frac{1}{2}\int \frac{ \dbar^3 k}{2 \omega(k)} \frac{ \dbar^3 p_{\bar{\chi}}}{2 E_{\bar{\chi}}}   \frac{1}{e^{\omega(k)/T}-1} (2 \pi)^4 \delta^{(4)}(K - p_\chi - p_{\bar{\chi}}) \sum_\mathrm{d.o.f.} \abs{\mathcal{M}}^2_{\gamma^* \rightarrow \chi \bar{\chi}}\eeq 
proceeds through direct computation. We find
\begin{align} C(p_\chi, t)_{\gamma^*_\ell \rightarrow \chi \bar{\chi}} &= \frac{ Q^2 e^2 }{4 \pi p_\chi} \int  \frac{dk\, \omega_\ell (k) Z_\ell(k) }{k\, (e^{\omega_\ell(k)/T}-1)}  \left(2 E_\chi (\omega_\ell(k) - E_\chi) - m_\ell(k)^2/2\right)
 \\ C(p_\chi, t)_{\gamma^*_t \rightarrow \chi \bar{\chi}} &= \frac{ Q^2 e^2 }{4 \pi p_\chi} \int \frac{dk\,k Z_t(k)}{\omega_t(k) (e^{\omega_t(k)/T}-1)}  \left(2 p_\chi^2 - \frac{( 2 E_\chi \omega_t(k) - m_t(k)^2)^2}{2 k^2} +m_t(k)^2\right)
\end{align}
where the limits of the $k$ integral are determined by the requirement that $x_0 = (2 E_\chi \omega_{\ell, t}(k) - m_{\ell, t}(k)^2)/2 k p_\chi$ lies in the range $[-1,1]$. The limits of integration cannot be solved for in closed form because of the nontrivial dispersion relations, so the phase space must be determined numerically.

\begin{figure*}[t]
\includegraphics[width=\textwidth]{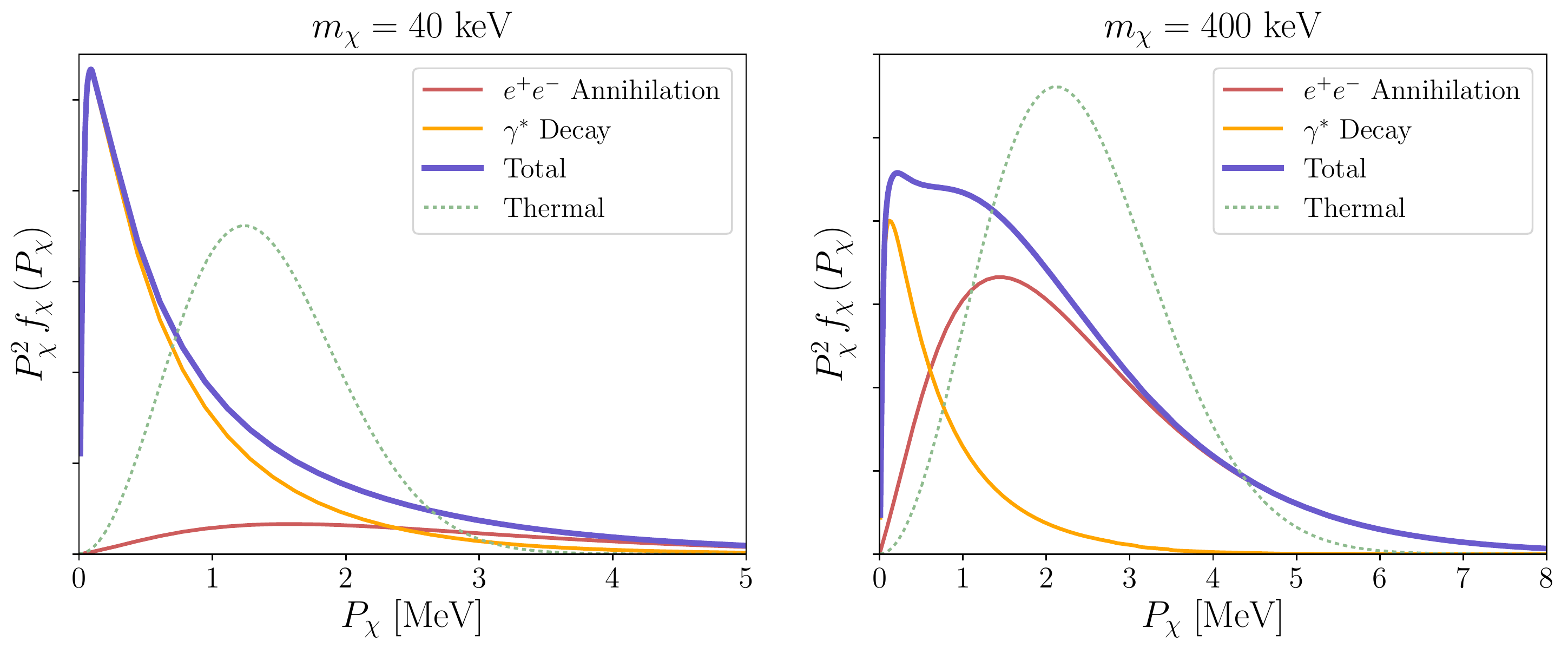}
\caption{A comparison of the contributions to the phase space for $m_\chi = 40$~keV (left) and $m_\chi = 400$~keV (right). The momenta shown here are comoving, $P_\chi \equiv a p_\chi$ where $a=1$ corresponds to $T= 1$~MeV. The phase space is normalized to the comoving DM relic abundance for each mass depicted. The plasmon contribution dominates more at low masses than at high masses because freeze-in through this channel persists for longer at lower masses, ending when the plasmon mass is at threshold, $\omega_p \sim 2 m_\chi$. Also shown (dashed lines) are the phase space distributions that would arise if the DM could thermalize within its own sector, conserving $\left<P_\chi^2\right>$ for non-relativistic DM.}
\label{masscomp_phase}
\end{figure*}

\begin{figure}[t]
\includegraphics[width=0.7\textwidth]{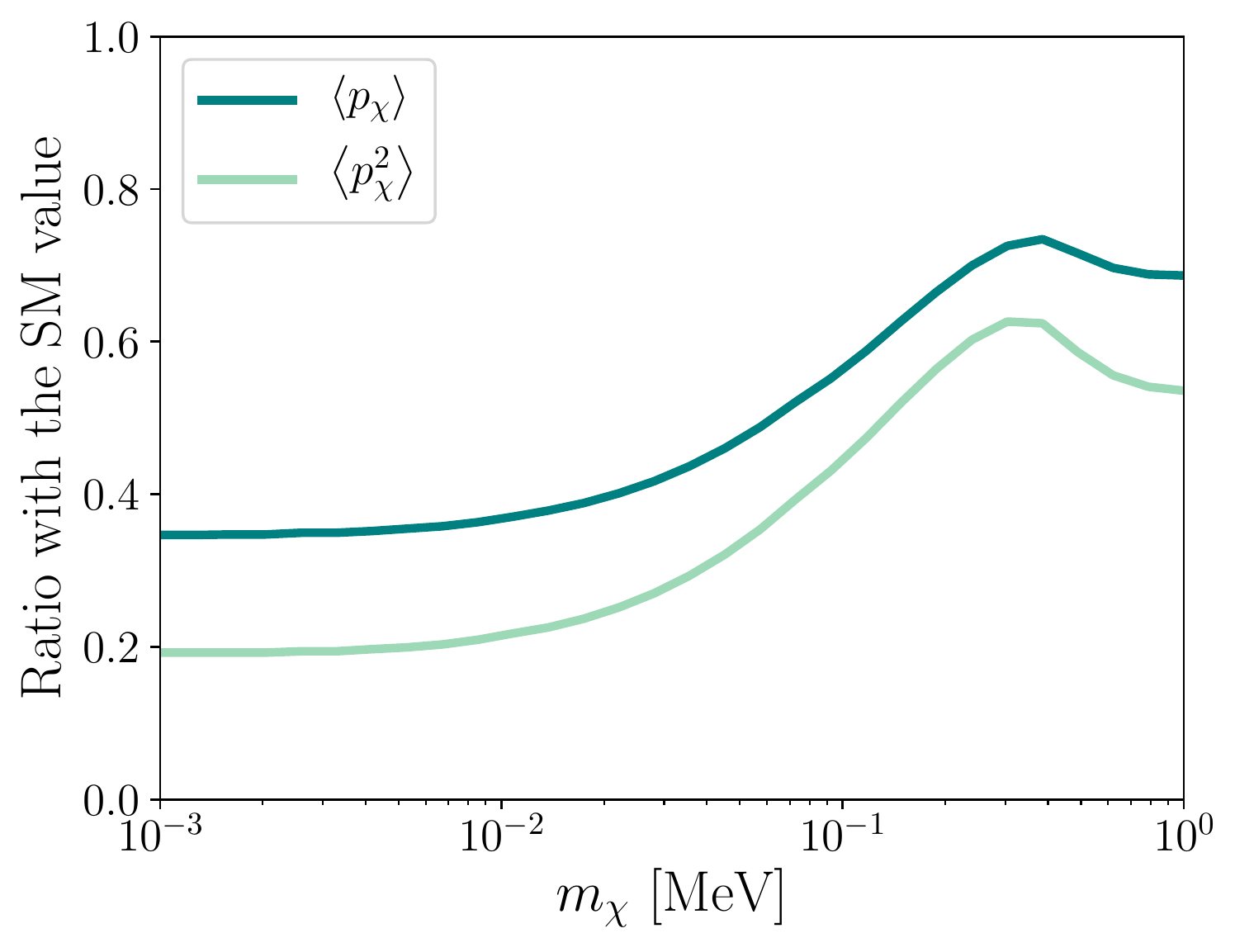}
\caption{A comparison between moments of the DM phase space and the SM photon phase space as a function of DM mass. For reference, the moments for the SM photon are $\langle p_\gamma \rangle = 2.7\, T_\gamma$ and $\langle p_\gamma^2 \rangle = 10.35\, T_\gamma^2$. While the DM phase space is not thermal, these moments can be thought of as relating to the DM effective temperature, which will have ramifications for the subsequent cosmology. As the DM mass rises, the effective temperature increases because $e^+ e^-$ annihilations become more important than plasmon decays and have a comparatively fatter high-$p_\chi$ tail. At even larger masses where $m_\chi$ is comparable to $m_e$, that high-$p_\chi$ tail is suppressed because the DM mass becomes relevant to the kinematics of annihilation, causing the effective temperature to drop. }
\label{Teff}
\end{figure}

The evolution of the phase space from plasmon decays is shown in the right panel of Fig.~\ref{freezephase}, and our results for the combined phase space can be found in Fig.~\ref{masscomp_phase}. The distributions are noticeably nonthermal due to plasmon decays. Fig.~\ref{Teff} compares the average momentum and momentum-squared of the DM to the SM photons, which serves as a useful metric to determine the DM free-streaming and suppression of the growth of structure.

\onecolumngrid

\subsection{Effect of DM-SM scattering \label{sec:scattering}}

We argue here that the effect of DM-SM scattering on the DM phase-space distribution is small from freeze-in until the onset of recombination. The relevant quantity is the momentum-transfer rate, which we estimate in the limits where the DM is relativistic and non-relativistic. We do not consider scattering by relativistic, charged SM particles because this is only relevant for electrons during freeze-in; during freeze-in, the number density of DM is many orders of magnitude smaller than the number density of electrons and the effect of electron-DM scattering is suppressed by $n_\chi/n_e$ relative to the dominant effect of electron-positron annihilations on the phase space. 
As outlined below, DM-SM scattering becomes more important at low velocities, corresponding to later cosmological times. This can affect CMB anisotropies and the cosmological 21~cm signal, and we provide more detailed calculations in that context in our companion paper~\cite{inprep}.

In the limit of relativistic DM scattering with non-relativistic SM particles (the case after freeze-in until $T_\gamma \sim m_\chi$), the differential cross section with respect to the center-of-mass scattering angle $\theta_\text{CM}$ is given by
\begin{align}
     \frac{d\sigma_{\chi b}}{d\cos \theta_\text{CM}} = \frac{ \pi   Q^2 \alpha^2 }{ p_\text{CM}^2} \frac{  (1+ \cos \theta_\text{CM})}{(1 - \cos \theta_\text{CM} +m_D^2/2p_{\text{CM}}^2)^2},  
     \label{eq:scattering_relativistic}
\end{align}
where $p_{\text{CM}}\equiv | \vec{p}_{\text{CM}}|$ is the momentum in the CM frame. Here we have taken $p_\chi \ll m_e$, which is a good approximation after freeze-in has ended. In this approximation, the dependence on the SM particle mass drops out, making scattering with electrons and protons equally important (we refer to them collectively as ``baryons,'' in the remainder of this discussion, hence the subscript $b$ in the cross section). The dependence on the Debye mass $m_D$ comes from the photon propagator for electric scattering in a medium~\cite{Blaizot:1995kg}. The usual $t$-channel divergence is thus regulated in the forward-scattering limit by the Debye angle, defined as $\theta_D \equiv m_D/p_{\text{CM}}$. Once the plasma has become non-relativistic with $T_\gamma \lesssim m_e$,  the Debye mass is given by \beq m_D = \sqrt{4 \pi \alpha n_e/T_\gamma} = 3.7\times 10^{-6}\, T_\gamma  \eeq
in natural units, assuming $\Omega_b h^2 = 0.022$ \cite{Aghanim:2018eyx} and that the ionization fraction is unity. The momentum transfer cross section is defined for DM self-scattering in Eq.~\eqref{eq:transfer} and the analogous definition applies for scattering between DM and SM particles. For relativistic DM, we find that in the limit of the Debye angle $\theta_D \ll 1$
\begin{align}
    \sigma_{T,\, \chi b} = \frac{4 \pi Q^2 \alpha^2}{p_\chi^2}  \log  \frac{2}{\theta_{D}} .
\end{align}
Since $m_b \gg m_\chi$ and the baryons are non-relativistic, the DM momentum in the CM frame can be approximated by the DM momentum in the comoving frame, $p_\chi$. As illustrated in Fig.~\ref{Teff}, the typical DM momentum is comparable to the SM photon temperature, with both quantities redshifting after freeze-in. Therefore, we can estimate the momentum transfer rate per DM particle and per Hubble time as
\begin{align}
   \frac{ n_p \sigma_{T, \,\chi b}}{H} \approx 5.3 \times 10^{-11} \, \left( \frac{ Q}{10^{-10}} \right)^2 \left( \frac{\MeV}{T_\gamma} \right)  ,
\end{align}
where $n_p \approx 1.5 \times 10^{-10}\, T_\gamma^3$ and $p_\chi \approx 0.4 \, p_\gamma \approx T_\gamma$. For $T_\gamma$ in the keV-MeV range and $Q < 10^{-10}$ for freeze-in, this rate is tiny and thus scattering in this regime has a negligible effect on the DM phase space. 

For scattering of non-relativistic DM and charged SM particles, the differential cross section is instead given by \begin{align}
     \frac{d\sigma_{\chi b}}{d\cos \theta_\text{CM}} = \frac{2 \pi   Q^2 \alpha^2 }{ \mu_{\chi b}^2 v^4} \frac{  1}{(1 - \cos \theta_\text{CM} +m_D^2/2p_{\rm CM}^2)^2},  
     \label{eq:scattering_NR}
\end{align}
where $\mu_{\chi b}$ is the  reduced mass of the DM and baryon, $\mu_{\chi b} = m_\chi m_b/(m_\chi+m_b)$, $v$ is the relative velocity between DM and SM particles,  and $ p_{\rm CM} = \mu_{\chi b} v$. The momentum transfer cross section is
\begin{align}
	\sigma_{T,\, \chi b} =  \frac{4 \pi \,  Q^2 \alpha^2 }{ \mu_{\chi b}^2 v^4} \log \frac{2}{\theta_{D}}
    \label{eq:NRmomentumtransfer} \, ,
\end{align}
where again we take the $\theta_D \ll 1$ limit.
Note that the Coulomb logarithm appearing here differs from the one that appears in the often-quoted Ref.~\cite{McDermott:2010pa}; however, that reference did not include the Debye mass in the photon propagator, as discussed in Appendix~\ref{appendix:debeye}. Compared to the Coulomb logarithm in Ref.~\cite{McDermott:2010pa}, our treatment of the Debye mass results in a factor of $2.5-3$ smaller momentum transfer rate at recombination; this will translate to a weaker CMB bound on generic millicharged DM than has been reported previously~\cite{Dvorkin:2013cea,Xu:2018efh,Slatyer:2018aqg,Kovetz:2018zan,Boddy:2018wzy}, which we explore in more detail in our companion paper~\cite{inprep}.

Given the velocity scaling in Eq.~\eqref{eq:NRmomentumtransfer}, momentum transfer is most important at late times. For freeze-in couplings, there may be a substantial effect at the recombination epoch. In particular, momentum transfer during this epoch leads to a drag force between the DM and baryon fluids, which can affect CMB anisotropies~\cite{Dubovsky:2001tr,Dvorkin:2013cea,Boddy:2018kfv,Boddy:2018wzy}. The CMB bounds {\emph{require}} that the momentum transfer rate is slow compared to the rate of Hubble expansion at $z \approx 1100$, thus limiting the possible effect on the DM phase space. We calculate the bounds in detail in the companion paper~\cite{inprep}, properly accounting for the velocity distribution for freeze-in DM with the updated Coulomb logarithm.

In addition to DM-baryon scattering as discussed above, DM-photon scattering is possible. However, these processes do not have the low-velocity $v^{-4}$ enhancement in the rate and the cross section scales as $Q^4$, so the effects are negligible. In the model with a dark photon $A'$, scattering processes such as  $ e^-+ \gamma \to  e^-  + A'$ are also possible and scale only as kinetic mixing squared $\kappa^2$. However, these processes are still negligible compared to DM-baryon scattering since they lack the low-$v$ enhancement and have an additional large suppression due to the in-medium kinetic mixing effects, as discussed in Section~\ref{sec:darkphoton}. Processes like $\chi + \gamma\to \chi  + A'$ scale as $Q^2 g_\chi^2$; these also lack the $v^{-4}$ enhancement and any enhancement (relative to DM-baryon scattering) from the large photon-to-baryon ratio is more than compensated by the factor of $g_\chi^2$, even at the largest values of $g_\chi$ that saturate SIDM bounds. 

\subsection{Effect of DM-DM scattering \label{sec:DMscattering}}

In the absence of a dark photon, DM self scattering is proportional to $Q^4$, rendering it entirely negligible. However, self-interactions of the DM can effectively thermalize the phase space distribution in the model with a dark photon. The rate for dark photon mediated DM scattering is proportional to $g_\chi^4$, and thus may be important if $g_\chi$ is sufficiently large compared to $\kappa$. Similar to DM-baryon scattering, the cross section scales as $1/v^4$ and so these effects are most important at later times when the DM is cooler. Sufficient levels of self-scattering will convert a free-streaming phase space distribution into a Maxwell-Boltzmann or Gaussian velocity distribution. In the non-relativistic limit, the quantity $\langle a(t)^2 p_\chi^2 \rangle$ will remain the same after this process (by conservation of comoving energy), although other moments of the phase space differ.

To determine when self-scattering becomes important, we estimate the redshift $z_{\rm therm}$ when the momentum transfer rate per DM particle and per Hubble time is order unity:
\begin{align}
    \frac{ n_\chi \sigma_{T,\,\chi \chi}  v}{ H(z_{\rm therm})} = 1
\end{align}
where $v$ is the relative velocity between DM particles and $\sigma_{T,\, \chi \chi}$ is the self-scattering momentum transfer cross section given in Eq.~\eqref{eq:transfer}, with the dark photon mass regulating the forward scattering instead of the Debye mass that is present for DM-baryon scattering. Using the ratio of the average DM momentum to the photon momentum in Fig.~\ref{Teff}, we approximate the relative velocity as $v \approx p_\chi/m_\chi \approx  T_\gamma(z)/m_\chi$. In this estimate, we have assumed that DM is non-relativistic at the time self-interactions become important.

\begin{figure}
\centering
\includegraphics[width=0.7\textwidth]{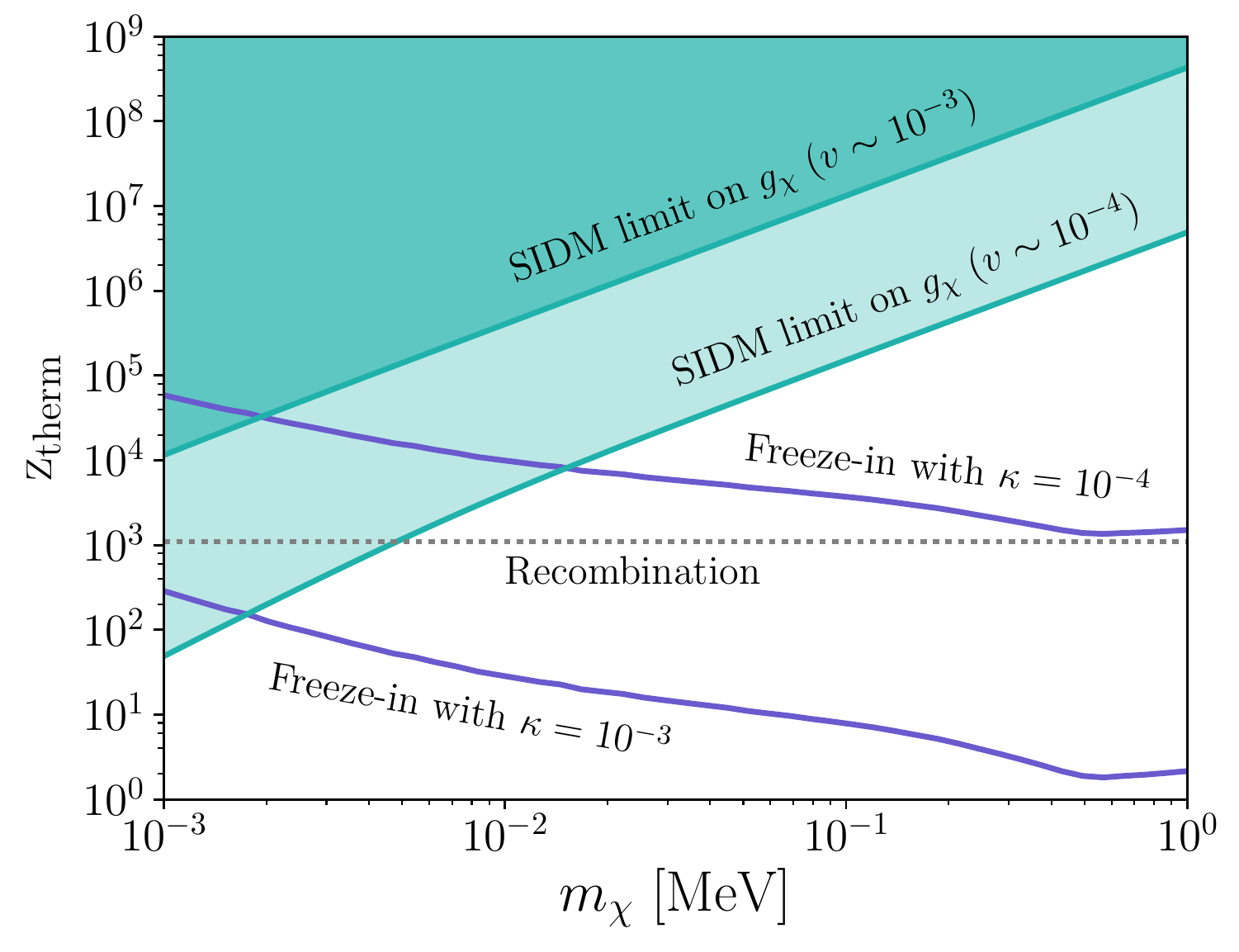}
\caption{The approximate redshift when DM self-scattering becomes important, $z_{\rm therm}$, as a function of DM mass in the model with dark photon mediated interactions. The freeze-in relic abundance is determined by $Q= g_\chi \kappa /e$ and we show $z_\text{therm}$ assuming two values of $\kappa$ (where $g_\chi$ is fixed to obtain the DM relic abundance). The epoch when DM self-thermalization becomes relevant is highly sensitive to the choice of couplings, which can yield different results for CMB observables depending on whether thermalization occurs before recombination. Note that DM halo formation is neglected in this estimate. Also shown are bounds on DM self-thermalization which come from the SIDM limits on $g_\chi$ in Eq.~\eqref{SIDMbound}. For illustration, we assume $\sigma_{T,\,\chi \chi} \lesssim 1$~cm$^2/$g for scattering via an ultralight mediator and show both $v\sim10^{-3}$ and $v\sim10^{-4}$, speeds relevant to a halo the size of the Milky Way and to a dwarf galaxy. In this figure we have taken $m_{A'} = 10^{-14}$~eV, which is sufficiently light that the constraints on the kinetic mixing parameter $\kappa$ are rather weak. \label{fig:DMtherm} }
\end{figure}

The self scattering randomizes the DM velocities while preserving the average kinetic energy $ \tfrac{3}{2} T^{\rm eff}_\chi(z) \equiv \langle p_\chi^2 \rangle/(2 m_\chi)$, where $p_\chi$ is physical momentum and the average momentum-squared is given in Fig.~\ref{Teff}. After self-scattering becomes significant, the DM phase space is described by a thermal Maxwell-Boltzmann distribution,
\begin{align}
	f_{\rm DM}(p_\chi,z) = n_{\rm DM}(z) \, \left( \frac{2\pi }{ m_\chi T_\chi^{\rm eff}(z)} \right)^{3/2} 4\pi p_\chi^2 \exp \left( - \frac{p_\chi^2 }{2 m_\chi T_\chi^{\rm eff}(z) } \right) ,
    \label{eq:fdm_PSD_Gaussian}
\end{align}
where $n_{\rm DM}(z)$ is the DM number density.

Fig.~\ref{fig:DMtherm} shows the redshift of thermalization for two representative choices of $\kappa$ (thus fixing $g_\chi$ to yield the observed relic abundance), where we see the assumption of non-relativistic DM is a reasonably good approximation in our estimates. Since the phase space calculations here will be an input to determining CMB constraints on freeze-in DM, we compare $z_{\rm therm}$ with the redshift of recombination $z \approx 1100$. For constraints from structure formation, a range of redshifts will be relevant. We also show some fiducial limits from SIDM, which give upper bounds on $g_\chi$. Fig.~\ref{fig:DMtherm} illustrates that the DM phase space at the time of recombination depends sensitively on the model parameters and on the robustness of SIDM limits in different astrophysical systems. For the largest values of $g_\chi$ consistent with the weaker assumed SIDM bounds, the DM phase space is described by a Maxwell-Boltzmann distribution at the time of recombination for all the DM masses we consider. However, for $\kappa = 10^{-3}$ (which is consistent with bounds on ultralight dark photons), $g_\chi$ is small enough that DM self-interactions are not important at recombination and the phase space is described by the results of Sections~\ref{sec:ann_phasespace}-\ref{sec:plasmon_phasespace}. The comparison of the free-streaming and thermalized phase space can be seen in Fig.~\ref{masscomp_phase}.

\section{Results and Discussion}
\label{summary}
\begin{figure}
\centering
\includegraphics[width=0.7\textwidth]{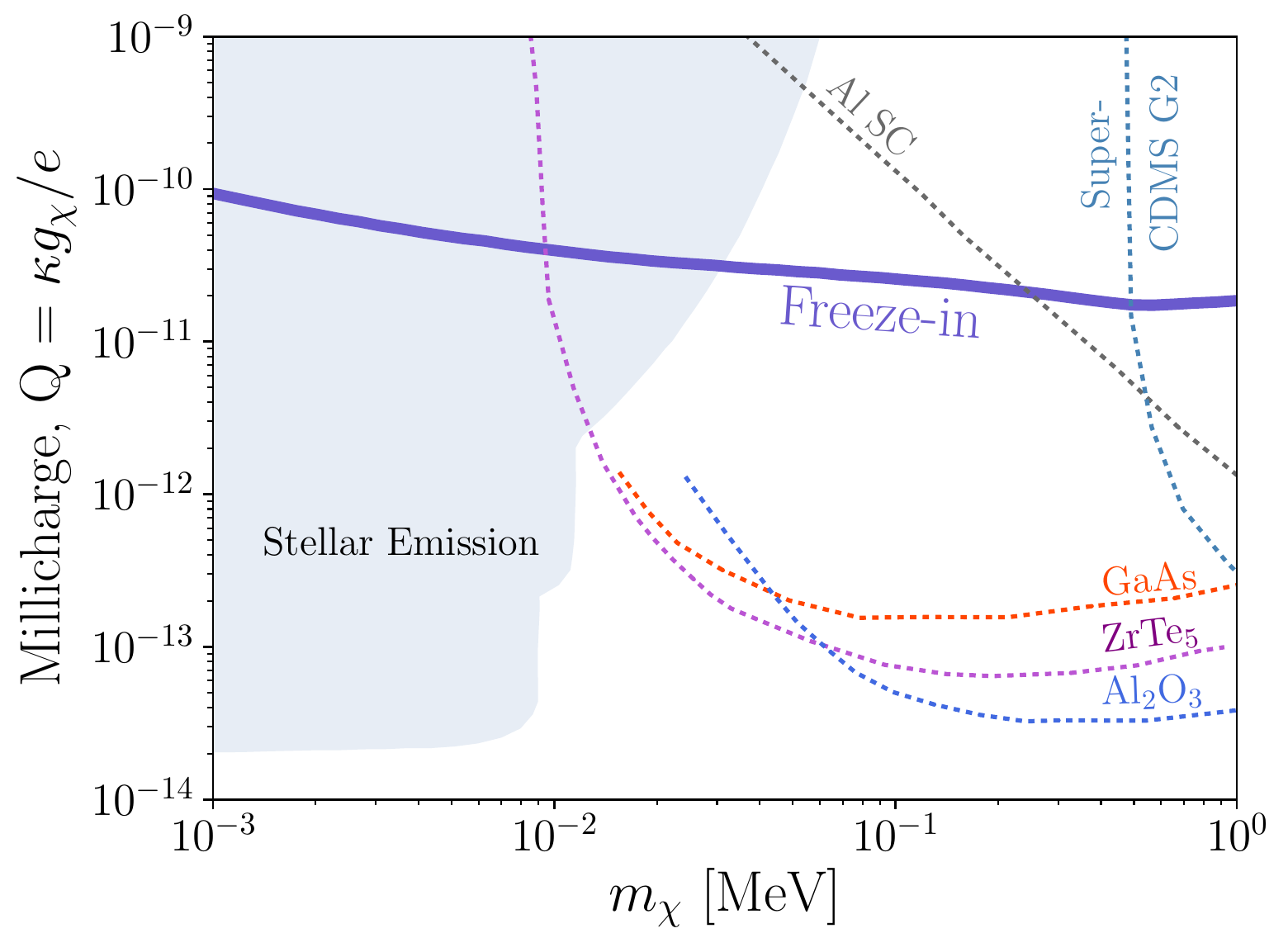} 
\caption{Summary plot including early-universe plasma effects for the parameter space of sub-MeV freeze-in DM. The correct DM relic abundance is obtained for couplings on the freeze-in line. We show constraints coming from emission of DM pairs in white dwarf, horizontal branch and red giant stars~\cite{Vogel:2013raa}, while bounds from emission of DM pairs in supernovae apply for $Q \gtrsim 10^{-9}$~\cite{Chang:2018rso}. Dotted lines are projected sensitivities of proposed direct detection experiments as in Fig.~\ref{DDshift}.  
\label{fig:summaryplot}}
\end{figure}

In this paper, we have shown that DM freeze-in through a light vector mediator is substantially affected by plasmon decay, which constitutes a new production channel. This is an efficient way of producing sub-MeV DM and is dominant over SM fermion annihilation for masses below a few hundred keV. To account for this extra production channel, the couplings between the DM and the SM must be reduced in order to obtain the observed relic abundance of DM. For the lightest DM masses that are accessible to low-threshold direct detection experiments, the predicted cross section is lowered by roughly an order of magnitude. Updated predictions for freeze-in through a light vector mediator are shown in Fig.~\ref{fig:summaryplot}.

The presence of this channel also affects the DM phase space. In the absence of plasmon decays, the DM is never technically thermal but it acquires a distribution that appears thermal by inheriting the electron phase space distribution at the time of production. At early times $f_{\chi, \, e^+e^-}(p_\chi)\sim e^{-p_\chi/T_{\chi, \, e^+e^-}}$, where $T_{\chi,\, e^+e^-}$ is an effective DM temperature inherited from the electrons; at late times, this exponential distribution persists because the DM does \emph{not} thermalize to give the Maxwell-Boltzmann distribution that would be expected for non-relativistic matter in equilibrium. On the other hand, the plasmon decay channel yields a DM phase space distribution that never appears thermal, which can be attributed to the running of the plasmon mass with temperature and the fact that plasmon decays occur dominantly as the plasmon wavenumber $k\rightarrow 0$. For DM masses where plasmon decays are the dominant production mode, the phase space is peaked at low momentum and has a long tail; for DM masses where contributions from both channels are important, the phase space distribution is bimodal.

Though the DM is born with a highly non-thermal distribution, it may be possible for the DM to thermalize with itself under the right circumstances. For DM that is only charged under the SM $U(1)_{EM}$ with millicharge $Q$, the thermalization rate is  suppressed by a factor of $Q^4$ where the requisite $Q$ to produce the DM relic abundance is $Q \sim \mathcal{O}\left(10^{-11}\right)$. If the DM is also charged under a dark $U(1)$ gauge group that kinetically mixes with the SM $U(1)_{EM}$ (with mixing parameter $\kappa$), it may be possible for DM self-scattering to thermalize the DM phase space distribution. In this case, $Q = \kappa g_\chi/e$ (where $\kappa$ can take on a wide range of values) and DM self-scattering via the dark photon scales as $g_\chi^4$, meaning that with the appropriate choice of $\kappa$ and $g_\chi$ it is possible to efficiently self-scatter while still producing the observed relic abundance. The coupling $g_\chi$ cannot be arbitrarily large due to observational limits on SIDM in astrophysical systems; however, there is a range of $g_\chi$ where self-scattering thermalizes the DM before recombination and where the SIDM bounds are simultaneously satisfied. Energy is conserved within the DM fluid, so for non-relativistic DM $\left<p_\chi^2\right>$ will be conserved and the resulting distribution has a well-defined notion of temperature.

Although the freeze-in DM phase space distribution may not be thermal, it is still informative to take moments of the distribution. When comparing the first and second moments of $f_\chi(p_\chi)$ to the equivalent quantities for the SM photon bath, we find that the typical DM momentum is similar to the typical photon momentum, $\langle p_\chi \rangle  \approx (0.4-0.7)\times  \langle p_\gamma \rangle$ depending on the DM mass. In other words, the DM is born considerably warmer than what is typically assumed for cold DM initial conditions. This will have ramifications for cosmology in two key ways:
\begin{itemize}
    \item Freeze-in DM will behave like warm DM, leading to suppression of the matter power spectrum below some physical scale roughly corresponding to the free-streaming length. This effect is not already captured by existing limits on warm DM, where different DM phase space distributions are assumed. To understand this suppression quantitatively, a Boltzmann code is necessary that accounts for the potentially nonthermal phase space from freeze-in. Having understood this, it will be possible to constrain DM freeze-in via a light vector mediator using probes of the matter power spectrum and the halo mass function.
    \item Existing CMB limits on DM with an effective millicharge do not straightforwardly apply to the case of freeze-in. These limits stem from a DM-baryon drag; because the drag is highly sensitive to the relative DM-baryon velocity (the cross section scales like $\sim v^{-4}$), modifications to the DM phase space can substantially alter the size of the effect. Existing limits have made the assumption of cold dark matter, and the larger DM velocities for freeze-in will lead to reduced drag force. Taking into account the updated Debye logarithm (which may weaken existing limits by a factor of $\sim 2-3$), the limit on freeze-in will be further reduced compared to previously reported results.
\end{itemize}
Both of these effects will be thoroughly explored in our companion paper~\cite{inprep}, which will place restrictions on the range of masses where DM freeze-in via a light mediator is observationally viable.

\section*{Acknowledgments}
We thank Masha Baryakhtar, Asher Berlin, Simon Knapen, Jung-Tsung Li, Adrian Liu, Aneesh Manohar, Sam McDermott, Julian Mu\~noz, and Tomer Volansky for helpful discussions. We acknowledge the importance of equity and inclusion in this work and are committed to advancing such principles in our scientific communities. CD was supported by NSF grant AST-1813694, Department of Energy (DOE) grant DE-SC0019018, and the Dean's Competitive Fund for Promising Scholarship at Harvard University. TL is supported by an Alfred P. Sloan Research Fellowship and Department of Energy (DOE) grant DE-SC0019195. Parts of this paper were prepared while at the KITP, supported by the National Science Foundation under Grant No. NSF PHY-1748958. KS is supported by a National Science Foundation Graduate Research Fellowship and a Hertz Foundation Fellowship. KS is grateful for the hospitality of the Lisanti group at Princeton University and of the Center for Cosmology and Particle Physics at New York University, where part of this work was completed.

\appendix

\section{Evolution of the SM bath}
\label{clock}
Throughout this work, we take the properties of the SM thermal bath to be given by their equilibrium values at zero chemical potential. The photons and neutrinos are relativistic gases with energy and entropy densities \beq 
\rho_\gamma = \frac{\pi^2 T^4}{15}\mathrm{,}\quad \quad
s_\gamma  = \frac{4 \rho_\gamma }{3 T}\mathrm{,}\quad \quad \rho_\nu = \frac{7 \pi^2T_\nu^4}{40} \mathrm{,}\quad \quad s_\nu = \frac{4 \rho_\nu}{3 T_\nu} .
\eeq
Here we distinguish between the neutrino and SM bath temperatures $T$ and $T_\nu$; in this work we assume that the neutrinos kinetically decouple at a temperature that is higher than relevant for sub-MeV freeze-in and that their temperature evolves adiabatically $T_\nu \sim 1/a$ during this epoch, which is a good approximation at the percent level. We also ignore the negligible neutrino masses. Meanwhile, the electrons are transitioning from being relativistic to being non-relativistic, so we use the unapproximated expressions for the energy and entropy density,
\beq \rho_e = \frac{2}{\pi^2} \int_{m_e}^\infty dE \,\frac{E^2 (E^2 - m_e^2)^{1/2}}{e^{E/T}+1}, \quad   p_e = \frac{2}{3 \pi^2} \int_{m_e}^\infty dE \,\frac{(E^2 - m_e^2)^{3/2}}{e^{E/T}+1},  \quad s_e = \frac{p_e+\rho_e}{T}. 
\eeq

\begin{figure}
\begin{center}
\includegraphics[width=0.5\textwidth]{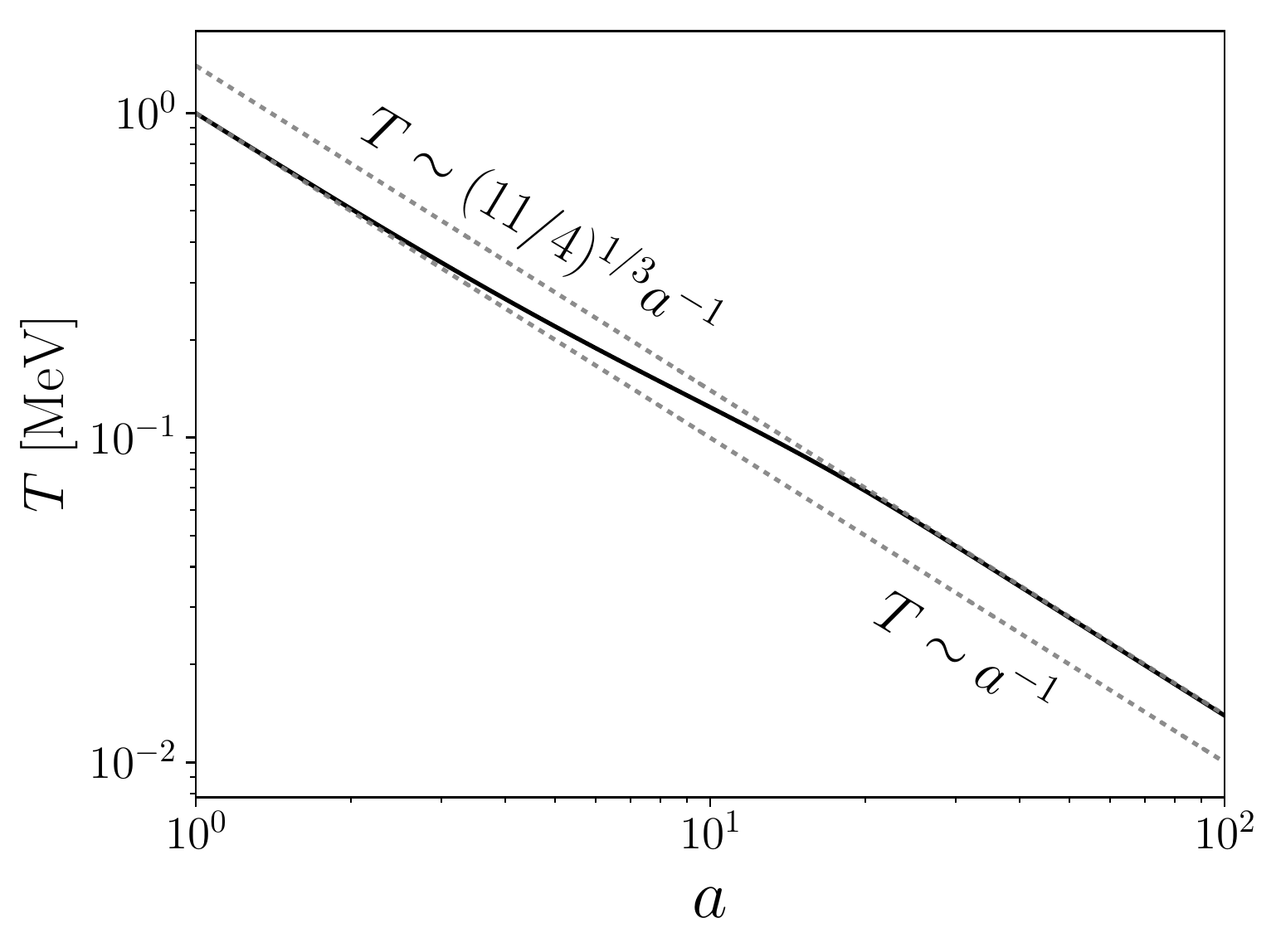}
\end{center}
\caption{The non-adiabatic temperature evolution of the SM thermal bath during freeze-in.}
\label{temp_evolve}
\end{figure}

Throughout the evolution of the SM bath, we require conservation of entropy. Since we are assuming adiabatic evolution of the neutrino temperature, its entropy $s_\nu(T_\nu) a(T)^3$ is constant by definition. The remaining constraint equation on the temperature evolution is then \beq \left( s_\gamma(T)+ s_e(T) \right) a^3= \mathrm{const.},\eeq
which yields a smooth temperature evolution $T(a)$, as shown in Fig.~\ref{temp_evolve}. After the electrons have fully left the bath, we recover the usual result $T_\nu=(4/11)^{1/3} T$. 
We can then use this temperature evolution to evolve the Hubble parameter smoothly through the transition as the electrons leave the thermal bath, 
\beq H^2(a) = \frac{\rho_e(T(a))+ \rho_\gamma(T(a)) + \rho_\nu(T_\nu(a))}{3 M_\mathrm{Pl}^2} \eeq 
with $M_\mathrm{Pl}$ the reduced Planck mass. Both the temperature and Hubble evolution feed into the calculations of the DM relic abundance and phase space in the main body of the text. 

\section{In-medium plasma properties}
\label{plasma}
In this Appendix, we follow the discussion of Ref.~\cite{Braaten:1993jw}, where the case of plasmons decaying to neutrinos was considered. The key approximation developed in that work was to evaluate thermal quantities at typical velocities, where thermal integrals have the most support. Specifically, the typical electron velocity is given by $v_* = \omega_1/\omega_p$, defined in terms of the first mode frequency and plasma frequency, 
\beq \omega_1^2 = \frac{4 \alpha}{\pi }\int dp \frac{p^2}{E} \left(\frac{5}{3} v^2 - v^4\right) f_e(E) \eeq 
\beq \omega_p^2 =  \frac{4 \alpha}{\pi }\int dp \frac{p^2}{E} \left(1 - \frac{1}{3}v^2\right) f_e(E) ,\eeq
where $f_e$ is the phase space density of electron-positron pairs.
Protons can also be included but their contribution is negligible because protons are heavy and thus slow to respond to electric fields, and also because their number density is much lower than that of the electrons at the relevant epochs.

The electromagnetic polarization tensor can be written as a thermal integral and expressed in terms of the longitudinal and transverse polarization functions, $\Pi_\ell$ and $\Pi_t$, as 
\begin{align*} \Pi^{\mu \nu}\big(\omega, \vec{k}\big) &= \left(1, \frac{\omega}{k} \hat{k}\right)^\mu \left(1, \frac{\omega}{k} \hat{k}\right)^\nu \Pi_\ell(\omega, k)\\
&+\left(\left(0, \vec{\epsilon}_+\right)^\mu \left(0, \vec{\epsilon}_+\right)^\nu + \left(0, \vec{\epsilon}_-\right)^\mu \left(0, \vec{\epsilon}_-\right)^\nu \right) \Pi_t(\omega, k) ,
\numberthis\end{align*}
where $\omega$ and $\vec{k}$ are the plasmon energy and wavevector, and where the vacuum transverse polarization vectors $\vec{\epsilon}_\pm$ are chosen to be orthogonal to the direction of propagation and normalized to unity. In terms of the quantities above, the polarization functions can be approximated as 
\beq \Pi_\ell(\omega, k) = \frac{3 \omega_p^2}{v_*^2 } \left( \frac{\omega}{2 v_* k} \ln \left( \frac{\omega + v_* k}{\omega - v_* k}\right) -1 \right)\eeq
\beq \Pi_t (\omega, k) =  \frac{3 \omega_p^2}{2 v_*^2 } \left(\frac{\omega^2}{k^2} - \frac{\omega(\omega^2 - v_*^2 k^2)}{2 v_* k^3}\ln \left( \frac{\omega + v_* k}{\omega - v_* k}\right) \right).\eeq
These approximations are accurate up to $\mathcal{O}(\alpha)$ and up to $\mathcal{O}(k^2)$ at small $k$ for all electron temperatures and densities.

The effective propagator can then be constructed; in Coulomb gauge, its nonzero components are 
\begin{align}
&D^{00}(\omega, \vec{k}) = \frac{1}{k^2 - \Pi_\ell (\omega, k)} \\
& D^{ij} (\omega, \vec{k}) = \frac{1}{\omega^2 - k^2 - \Pi_t (\omega, k)} \left( \delta^{ij} - \hat{k}^i \hat{k}^j\right).\end{align}
The poles in the propagator yield the renormalized longitudinal and transverse dispersion relations for on-shell plasmons,
\beq \omega_\ell(k)^2 = \frac{\omega_\ell(k)^2}{k^2}\Pi_\ell (\omega_\ell (k), k) \quad \quad  \omega_t(k)^2 = k^2 +\Pi_t (\omega_t(k), k), \label{dispersion} \eeq
while the residues of the poles are identified as a combination of dressed polarization four-vectors, $\tilde{\epsilon}^\mu(k) \tilde{\epsilon}^\nu (k)^*$, for the appropriate polarization. The dressed polarization vectors are given by
\beq \tilde{\epsilon}_L^\mu(k) = \frac{\omega_\ell(k)}{k} \sqrt{Z_\ell(k)}\left(1, \vec{0}\right)^\mu \quad \quad \tilde{\epsilon}_\pm^\mu(k) = \sqrt{Z_t(k)} \left(0, \vec{\epsilon}_\pm\right)^\mu.\eeq
Given the approximations for $\Pi_\ell$ and $\Pi_t$ and the dispersion relations, the residue functions can be written as 
\beq Z_\ell(k)  = \frac{2 (\omega_\ell(k)^2 - v_*^2 k^2)}{3 \omega_p^2 - (\omega_\ell(k)^2 - v_*^2 k^2)}\eeq
\beq Z_t(k) = \frac{2 \omega_t(k)^2 (\omega_t(k)^2 - v_*^2 k^2)}{3\omega_p^2 \omega_t(k)^2 + (\omega_t(k)^2 + k^2)(\omega_t(k)^2 - v_*^2 k^2) - 2 \omega_t(k)^2(\omega_t(k)^2 - k^2)}.\eeq


\section{Plasmon decays through a dark photon}
\label{basis}

In this Appendix, we show that plasmon decays in the millicharge basis (Eq.~\eqref{eq:inmedium_interactions}) are identical to decays in the basis where the dark photon has a coupling $e \kappa J_{\rm EM}^\mu A^\prime_\mu$. In a thermal plasma, this coupling generates an in-medium mixing term in the Lagrangian given by $\kappa A_\mu \Pi^{\mu \nu} A^\prime_\nu$ where $\Pi^{\mu \nu}$ is the electromagnetic polarization tensor.
The matrix element in the dark photon basis is then given by
\beq i \mathcal{M} = i \kappa g_\chi \tilde{\epsilon}^\mu(k) \Pi_{\mu \nu}\left(\omega, \vec{k}\right) D_{A'}^{\nu  \alpha}\left(\omega, \vec{k}\right)\bar{u}(p_\chi) \gamma_\alpha v(p_{\bar{\chi}})
 \equiv  i \kappa g_\chi \tilde{\epsilon}^\mu(k) \bar{u}(p_\chi) \gamma_\alpha v(p_{\bar{\chi}}) \, \Gamma^{\alpha}_\mu, 
\eeq
where  $D_{A'}^{\nu \alpha}$ is the dark photon propagator. Taking the $m_{A'} = 0$ limit and working in Coulomb gauge, the propagator is given by 
\beq D_{A'}^{\nu \alpha}\left(\omega, \vec{k}\right) = \frac{\big(1, \vec{0}\big)^\nu \big(1, \vec{0}\big)^\alpha }{k^2} + \frac{\left(0, \vec{\epsilon}_+\right)^\nu \left(0, \vec{\epsilon}_+\right)^\alpha +\left(0, \vec{\epsilon}_-\right)^\nu \left(0, \vec{\epsilon}_-\right)^\alpha }{\omega^2 - k^2}.\eeq
Here we are ignoring in-medium corrections on the dark photon propagator, which are suppressed by factors of $\kappa^2$. Contracting $D^{\nu \alpha}_{A'}$ with $\Pi_{\mu \nu}$ yields a vertex 
\begin{align*} \Gamma^{\alpha\mu}\left(\omega, \vec{k}\right) &=
 -\frac{\Pi_t(\omega, k) \left(\left(0, \vec{\epsilon}_+\right)^\mu \left(0, \vec{\epsilon}_+\right)^\alpha + \left(0, \vec{\epsilon}_-\right)^\mu \left(0, \vec{\epsilon}_-\right)^\alpha\right)}{\omega^2 - k^2} + \frac{\Pi_\ell(\omega, k) \left(1, \frac{\omega}{k} \hat{k}\right)^\mu\big(1, \vec{0}\big)^\alpha}{k^2} \quad \quad \\
& = -\left(0, \vec{\epsilon}_+\right)^\mu \left(0, \vec{\epsilon}_+\right)^\alpha - \left(0, \vec{\epsilon}_-\right)^\mu \left(0, \vec{\epsilon}_-\right)^\alpha  +\left(1, \frac{\omega}{k} \hat{k}\right)^\mu\big(1, \vec{0}\big)^\alpha.\numberthis 
\end{align*}
In the second line, we have assumed on-shell transverse and longitudinal modes for the respective pieces of the vertex function and used the dispersion relations of Eq.~\eqref{dispersion}.
Contracting this with a dressed polarization vector for the external photon yields 
\begin{align}
\tilde{\epsilon}_L^\mu(k)  \Gamma^{\alpha}_\mu\left(\omega_\ell, \vec{k}\right) &= \frac{\omega_\ell(k)}{k} \sqrt{Z_\ell(k)}\left(1, \vec{0}\right)^\alpha\\
\tilde{\epsilon}_\pm^\mu(\vec{k})  \Gamma^{\alpha}_\mu \left(\omega_t, \vec{k}\right) &=  \sqrt{Z_t(k)} \left(0, \vec{\epsilon}_\pm\right)^\alpha,
\end{align}
which gives the same result as the vertex obtained in the millicharge basis. 

\section{Regulating forward scattering}
\label{appendix:debeye}
The differential DM-baryon scattering cross section can be written with respect to the CM angle $\theta_\text{CM}$ as \beq \frac{d \sigma}{d \cos \theta_\text{CM}} = \frac{\abs{\mathcal{M}}^2}{32 \pi s} .\eeq
In the limit where all of the particles are non-relativistic and where $m_{A'} \ll m_D$ (if a dark photon is even present in the theory), the matrix element  squared for DM-baryon Coulomb scattering is given by\footnote{Note that if the dark photon mass becomes large enough that it poses a relevant scale in the problem, then an additional factor of $q^4 / (q^2-m_{A'}^2)^2$ appears to account for the in-medium couplings in Eq.~\eqref{eq:inmedium_L}.} 
\beq \abs{\mathcal{M}}^2 \approx \frac{16 Q^2 e^4 m_\chi^2 m_b^2}{\left(q^2 - m_D^2\right)^2}  = \frac{4 Q^2 e^4 m_\chi^2 m_b^2}{p_\text{CM}^4 \left(\cos\theta_\text{CM} -1 - m_D^2/2 p_\text{CM}^2\right)^2} , \eeq 
where we averaged over initial spins and summed over final spins. Here $q$ is the momentum transfer four-vector which satisfies $q^2 = - 2 p_\text{CM}^2 (1-\cos \theta_\text{CM})$ in the CM frame, $p_\text{CM} = |\vec p_\text{CM}|$ is the magnitude of the 3-momentum in this frame, and $m_D$ is the Debye mass. This effective mass arises from considering the longitudinal polarization tensor of the plasma $\Pi^{0 0}$ with the appropriate photon kinematics ($\omega \ll |\vec q|$)~\cite{Blaizot:1995kg}, which corresponds to screened Coulomb scattering. It can also be understood as the effective mass appearing in the screened electric potential, which takes the form of a Yukawa potential~\cite{Blaizot:1995kg, braaten1991calculation,raffelt1986astrophysical} or as a scale appearing in the electric form factor for a thermal Gibbs ensemble of charged particles in the plasma~\cite{raffelt1986astrophysical}. Note that the transverse polarization tensor $\Pi^{i j }$, which corresponds to the magnetic scattering mode, vanishes in the static $\omega \ll |\vec q|$ limit~\cite{Blaizot:1995kg};
however, this mode of scattering is negligible for a non-relativistic plasma where its contribution is suppressed by factors of $v$~\cite{baym1990transverse,braaten1991calculation}. 

The Debye mass automatically regulates the forward scattering divergence in the transfer cross section
\begin{align*} \sigma_{T,\, \chi b} = \int d \cos \theta_\text{CM} \frac{d \sigma}{d \cos \theta_\text{CM}} (1- \cos \theta_\text{CM})
\approx \frac{4 \pi Q^2 \alpha^2 }{\mu_{\chi b}^2 v^4} \ln\left(\frac{2 p_\text{CM}}{m_D}\right), \numberthis \label{debyetheta}
\end{align*}
where in the second equality we have taken the approximation $s = (m_b+m_\chi)^2$ for non-relativistic particles and have also taken the approximation $p_\text{CM} \gg m_D$. In the CM frame $p_\text{CM} = \mu_{\chi b} v$ where $\mu_{\chi b} = m_b m_\chi /(m_b + m_\chi)$ is the DM-baryon reduced mass and $v$ is the relative velocity. If we had cut the integral by hand at some angle $\theta_D$ (rather than including the Debye mass in the propagator) we would have obtained a logarithm $\ln 2/ \theta_D$ so we identify the correct Debye angle as $m_D/p_\text{CM}$. This Coulomb logarithm also agrees with other DM-baryon scattering rates found in the literature, for instance in Refs.~\cite{davidson2000updated,Dubovsky:2003yn,raffelt1986astrophysical,Vogel:2013raa}.

This procedure yields a different logarithm than Ref.~\cite{McDermott:2010pa}, which has been used for recent CMB constraints on millicharged DM. In that work, the angular integral was cut by using the relation between impact parameter and scattering angle for (electric) Coulomb scattering, and requiring that the impact parameter for pairwise DM-baryon scattering not exceed the Debye length $\lambda_D = 1/m_D$. This translated to a minimum scattering angle that depended on the DM millicharge, with $\theta_{\rm min} = 2 Q \alpha/(3 T \lambda_D)$. The corresponding minimum momentum transfer in that case would be $|\vec q|^2 = 4 Q^2 \alpha^2 p_{\rm CM}^2 m_D^2/(9 T^2)$. For freeze-in where $p_{\rm CM} \approx T$ and $Q < 10^{-10}$, we see that $|\vec q|^2 \ll m_D^2$ and so we expect that the Yukawa-like form of the effective potential leads to a strong screening effect for modes of such large spatial size. In other words, the requirement of  Ref.~\cite{McDermott:2010pa} may not be restrictive enough because DM-baryon scattering is suppressed by factors of $Q$ relative to the strong collective effects in the plasma that give rise to the Debye mass. Because forward scattering is so peaked, the resulting transfer cross section is highly sensitive to the limits of integration and their procedure yields a transfer cross section that is a factor of $\sim 2-3$ larger than the one obtained with the procedure of Eq.~\eqref{debyetheta}. As a result, CMB limits on millicharged DM that use this result may be too strong.

\bibliography{freezein_particle}

\end{document}